\documentclass[ a4paper, 
                amsfonts, 
                amssymb, 
                amsmath, 
                reprint, 
                showkeys, 
                nofootinbib,
                superscriptaddress, 
                onecolumn, 
                notitlepage,
                aps,
                pra,
                longbibliography]{revtex4-1}

\usepackage{amsmath}
\usepackage[T1]{fontenc}    
\usepackage[utf8]{inputenc}  
\usepackage[top=20mm, bottom=20mm, outer=20mm, inner=20mm, marginparwidth=33mm, marginparsep=3mm]{geometry}
\usepackage{graphicx}
\usepackage[colorlinks=true, allcolors=blue]{hyperref}
\usepackage{physics}
\usepackage{tikz}
\usepackage{cleveref}
\usepackage{xcolor}
\usepackage[english]{babel}
\usepackage{soul}
\usepackage{etoolbox}
\usepackage{subcaption}
\usepackage{marginnote} 
\usepackage{placeins}
\usepackage[normalem]{ulem}

\definecolor{poissonSteve}{RGB}{255, 120, 80}

\newcommand{\comment}[2]{\marginnote{\textcolor{poissonSteve}{#1}} \textcolor{poissonSteve}{#2}} 

\robustify{\comment}

\crefname{equation}{Eq.}{Eqns.}
\crefname{figure}{Fig.}{Figs.}

\DeclareMathOperator*{\argmax}{arg \ max} 

\newcommand{\appsection}[2]{
  \item \hyperref[#1]{#2 \dotfill \pageref{#1}}
}

\newcommand{\appsubsection}[2]{
  \item \hspace*{1.5em}\hyperref[#1]{#2 \dotfill \pageref{#1}}
}
\makeatletter
\newcommand{\appendixfigures}{
  \setcounter{figure}{0}
  \renewcommand{\thefigure}{A\arabic{figure}}
}
\makeatother
\makeatletter
\newcommand{\appendixtables}{
  \setcounter{table}{0}
  \renewcommand{\thetable}{A\arabic{table}}
}
\makeatother
\makeatletter
\newcommand{\appendixsubsectionformat}{
  \renewcommand{\thesubsection}{\thesection.\arabic{subsection}}
  \@addtoreset{subsection}{section}
}
\makeatother

\begin{document}

\title{Adaptive Reconstruction of Bosonic Quantum States}

\author{Vasilisa Usova}
\affiliation{Institute for Quantum Optics and Quantum Information, Austrian Academy of Sciences, 6020 Innsbruck, Austria}
\affiliation{Institute for Experimental Physics, University of Innsbruck,  6020 Innsbruck, Austria}
\author{Phila Rembold}
\affiliation{Atominstitut, TU Wien, Stadionallee 2, 1020 Vienna, Austria}
\author{Ian Yang}
\affiliation{PSI Center for Photon Science, 5232 Villigen PSI, Switzerland}
\author{Marco Rossignolo}
\affiliation{Qruise GmbH, 66113 Saarbr{\"u}cken, Germany}
\author{Simone Montangero}
\affiliation{Dipartimento di Fisica e Astronomia “G. Galilei”, Università degli Studi di Padova, 35131 Padua, Italy}
\affiliation{INFN Istituto Nazionale di Fisica Nucleare, Sezione di Padova, 35131 Padua, Italy}
\author{Samuele Tosatto}
\affiliation{University of Innsbruck, Department of Computer Science and Digital Science Center, 6020 Innsbruck, Austria}
\author{Gerhard Kirchmair}
\affiliation{Institute for Quantum Optics and Quantum Information, Austrian Academy of Sciences, 6020 Innsbruck, Austria}
\affiliation{Institute for Experimental Physics, University of Innsbruck, 6020 Innsbruck, Austria}

\begin{abstract}
Bosonic quantum systems provide a hardware-efficient platform for quantum information processing but remain challenging to characterise due to their large Hilbert space and the high measurement cost of state tomography. Existing approaches estimate the fidelity with respect to a single target state, making them unsuitable for applications in which physically equivalent states differ by phase space translations, rotations, or other  transformations. Here, we introduce an adaptive reconstruction technique that estimates the fidelity with respect to a family of bosonic states while reconstructing the underlying Wigner function from a small number of measurements. The method combines a physics-informed parametric model with Bayesian inference, bootstrap, and active learning to iteratively select the most informative phase space sampling points. We implement the approach on a circuit quantum electrodynamics platform and benchmark it on Schrödinger cat states with amplitudes $\alpha\in[1,3]$. The reconstruction yields reproducible fidelity estimates within a few minutes, remains robust to substantial displacements and rotations in phase space despite using a mismatched prior, and is sensitive to subtle state imperfections. We further compare the adaptive strategy with existing Wigner function sampling protocols experimentally, demonstrating the advantage of adaptive sampling for measurement-efficient fidelity estimation with respect to a family of cat states. Finally, we incorporate the reconstructed fidelity into the figure of merit used in a proof-of-principle closed-loop quantum optimal control experiment, demonstrating the applicability of the method to autonomous optimisation of bosonic quantum states.
\end{abstract}

\maketitle

\section{Introduction}

Bosonic modes, owing to their intrinsically large Hilbert space, provide a hardware-efficient paradigm for quantum information processing~\cite{Joshi_2021, Cai_2021}. In contrast to conventional discrete-variable~(DV) qubits, bosonic qubits encode quantum information in superpositions of infinite-dimensional continuous-variable~(CV) states of a quantum harmonic oscillator. Bosonic qubits have recently attracted increasing attention as they enable resource-efficient implementations of quantum error correction protocols by exploiting redundancy within a single mode rather than distributing it across multiple qubits~\cite{gottesman_encoding_2001, Sivak_break_even_2023}. Schrödinger cat states, in particular, provide an interesting candidate~\cite{cochrane_macroscopically_1999, chamberland_building_2022,guillaud_repetition_2019}.

However, this increased encoding capacity comes at the cost of challenging state characterisation. While fidelity provides a natural measure of distance between an implemented and a target state, determining it in a CV Hilbert space is inherently resource-intensive for complex states, typically requiring full reconstruction via the Wigner function and a large number of measurements. This gave rise to the development of optimal sampling strategies~\cite{sivak_model-free_2022, LyonNN} that enable estimation of fidelity with respect to a target state. The situation is further complicated by effective degeneracies not reflected in the fidelity, since trivial operations, such as global phase rotations or displacements in phase space, yield quantum states that are practically equivalent. These challenges call for reconstruction protocols that do not target a single fixed state but instead identify the optimal fidelity within a relevant family of states, thereby reducing measurement overhead while retaining accuracy.

\begin{figure}[h]
    \centering
    \includegraphics[]{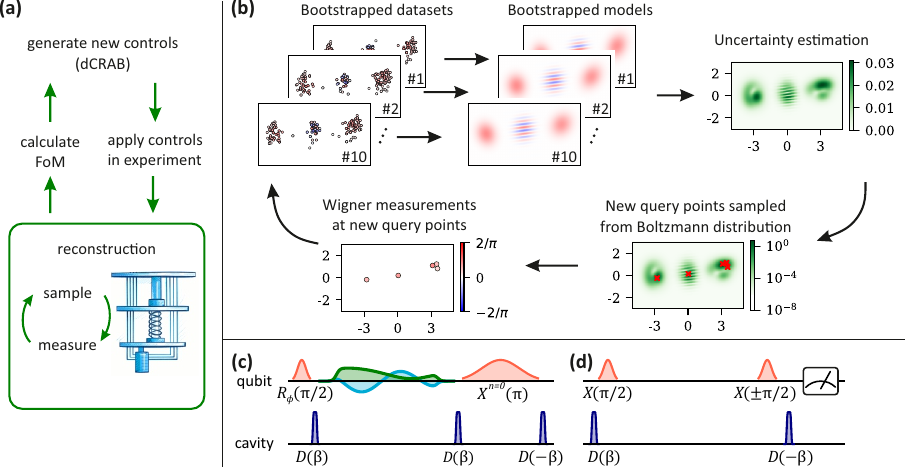}
    \caption{Closed-loop optimal control and bosonic state reconstruction. 
    \textbf{(a)} Closed-loop quantum optimal control in a cQED setup. A figure of merit (FoM) is calculated based on the result of the bosonic state reconstruction and passed to the optimisation algorithm that generates new control pulses for the next optimisation iteration. \textbf{(b)} Reconstruction algorithm workflow. Measured Wigner values are sampled with replacement to obtain 10 different datasets. These datasets are passed to 10 identically initialised models. As a result of maximum a posteriori (MAP) optimisation, models converge to a different set of parameters and produce various Wigner maps. Deviations between the generated Wigner maps are used to sample the new set of points to be measured (batch), shown as red crosses. \textbf{(c)} qcMAP pulse sequence. Qubit pulses are coloured in red, cavity pulses are coloured in violet. The optimiser applies corrections to the phase of the first pulse and to the qubit $I$ (green) and $Q$ (light blue) quadratures during the entangling operation, which is equivalent to waiting for $t = \pi/\chi$. \textbf{(d)} Wigner measurement sequence. Qubit pulses are coloured in red, cavity pulses are coloured in violet.}
    \label{fig:overview}
\end{figure}

Beyond these intrinsic challenges, state preparation protocols in realistic experimental setups are inevitably affected by higher-order nonlinearities in the Hamiltonian. When these terms are known, they can be accounted for in model-optimised (i.e., open-loop based) pulse sequences~\cite{QOC_review_1}. However, unaccounted Hamiltonian contributions, imperfect calibration, and system drifts may cause the actual dynamics to deviate from the model, thereby limiting the application of open-loop methods. These deviations highlight the need for adaptive, measurement-driven strategies that can be directly integrated into a direct feedback-based (i.e., closed-loop) quantum optimal control (QOC) framework to iteratively improve state fidelity and compensate for distortions on the experimental platform itself (Fig.~\ref{fig:overview}\textbf{(a)}).

In this work, we introduce a fast and accurate state reconstruction method for bosonic states, given a family of quantum states. The reconstructor adaptively distributes sampling points across the phase space (Fig.~\ref{fig:overview}\textbf{(b)}) according to the state family of interest, concentrating the measurements in the areas of the largest uncertainty. Our reconstruction technique yields a high-fidelity, reproducible Wigner map within a few minutes, requires no pretraining, and could be adapted to any CV platform using any bosonic state~\cite{saner_generating_2026}. 

We implement these ideas on a circuit quantum electrodynamics~(cQED) platform. Our setup combines a long-lived superconducting cavity with a qubit (see Appendix~\ref{app:ExpSetup}). The bosonic mode comprises a high-Q three-dimensional niobium cavity~\cite{heidler_Nb_cav_2021}, while the transmon qubit~\cite{TransmonKoch} provides the requisite nonlinearity and facilitates a wide array of cQED measurement and control techniques. The system is described in the dispersive regime by the Hamiltonian 
\begin{equation}
    \hat{H}/\hbar = \omega_{\mathrm{c}} \hat{c}^\dagger \hat{c} +\omega_{\mathrm{q}} \hat{q}^\dagger \hat{q} - \frac{K_\mathrm{q}}{2} \hat{q}^\dagger\hat{q}^\dagger \hat{q} \hat{q} - \frac{K_\mathrm{c}}{2} \hat{c}^\dagger\hat{c}^\dagger \hat{c} \hat{c} - \chi_{\mathrm{qc}} \hat{c}^\dagger \hat{c} \hat{q}^\dagger \hat{q},
    \label{eq:dispersive}
\end{equation}
where $\hat{c}^\dagger$ and $\hat{q}^\dagger$ are the creation operators for the cavity and qubit modes, respectively. The cavity and qubit have frequency $\omega_\mathrm{c/q}$ and nonlinearity $K_\mathrm{c/q}$ and interact with the dispersive interaction strength $\chi_\mathrm{qc}$. Using the dispersive interaction, we form cat states of the size $\alpha$ from the displaced coherent states $\ket{\alpha}$ with fringe phase $\varphi$ 
\begin{equation}
    |C_\alpha^\varphi\rangle = N_{\varphi} \left( |\alpha\rangle + e^{i\varphi} |-\alpha \rangle \right), \ N_{\varphi} = \left( 2 + 2e^{-2|\alpha|^2}\cos\varphi \right)^{-1}
\end{equation}
in a bosonic mode of the high-Q cavity with the qcMAP protocol~\cite{leghtas_Phys.Rev.A_87:4_2013}. We measure their Wigner function $W(\beta)$, where $\beta=I+iQ$ represents the phase space with conjugate variables $I$ and $Q$~\cite{vlastakis_Sci_342:6158_2013}, as shown in Fig.~\ref{fig:overview}\textbf{(c, d)}. 

As a proof of principle, we apply closed-loop, i.e. feedback-based, quantum optimal control (QOC)~\cite{QOC_review_1}, in which control pulses are iteratively refined based on feedback from experimental data (Fig.~\ref{fig:overview}\textbf{(a)}). The robustness of the closed-loop optimisation, as shown in discrete-variable systems~\cite{QOC_review_1}, is ensured in the continuous-variable case by the reproducibility of our reconstruction protocol. The objective is to improve the fidelity of a cat state within a family of cat states of fixed size and parity. We perform the optimisation by replacing the waiting time of the protocol (Fig.~\ref{fig:overview}\textbf{(c)}) with a qubit control pulse and varying the phase of the first pulse in the sequence, which does not require additional hardware resources. The feedback is provided by calculating a figure of merit (FoM) (Fig.~\ref{fig:overview}\textbf{(a)}), based on the state fidelity, obtained via the reconstruction algorithm (Fig.~\ref{fig:overview}\textbf{(b)}).

The paper is organised as follows. In Sec.~\ref{sec:Reconstruction}, we introduce the principles of our reconstruction technique and present its benchmarking results, concluding with a discussion of possible extensions to other classes of bosonic states and to alternate tomography techniques. In Sec.~\ref{sec:SamlingComparison}, we compare fidelity estimates obtained using various existing Wigner function sampling strategies with our adaptive approach. In Sec.~\ref{sec:ClosedLoop}, we demonstrate closed-loop optimisation of cat states using the reconstructed state fidelity as the FoM. Finally, we outline future directions toward more general closed-loop optimisation protocols in Sec.~\ref{sec:Conclusion}.

\section{Bosonic State Reconstruction} \label{sec:Reconstruction}

The goal of the reconstruction algorithm is to efficiently recover an accurate Wigner function of the underlying bosonic state within the shortest possible time. To integrate well with closed-loop optimisation protocols, the algorithm should meet the following criteria: (i) the fidelity estimation error must remain minimal; (ii) the fidelity estimate should be reproducible across trials; and (iii) the total reconstruction time must be constrained to a few minutes. To the best of our knowledge, no existing reconstruction techniques satisfy all of these requirements simultaneously. 

Bosonic state estimation is, in essence, an \textsl{active learning problem} \cite{settles_active_2012}, in that we have the freedom to select the measurement points $\beta$ that yield the most efficient estimate of the Wigner function. Bayesian experimental design \cite{chaloner_bayesian_1995,rainforth_modern_2024}, particularly \textsl{Bayesian active learning by disagreement} \cite{houlsby_bayesian_2011,gal_deep_2017} resolves such problems by iterating three core steps: (i) posterior parameters estimation (i.e. given a dataset, what is the most likely Wigner function?), (ii) epistemic uncertainty quantification (i.e. which measurement points — in our case phase space coordinate $\beta$ — yield highest uncertainty?), and (iii) sample selection (i.e. which coordinate $\beta$ should be used for the next measurement?). 

To balance computational constraints with sample efficiency, our approach maps to these steps as follows. First, we fit a parametrised physics-informed model of the  Wigner distribution to ensure precise fidelity estimation. Second, we combine Maximum A Posteriori (MAP) estimation — incorporating a Wishart prior to constrain parameters to realistic regimes — with bootstrap~\cite{Bootstrap,newton_approximate_1994}. Note that our bootstrap approach efficiently approximates epistemic uncertainty while drastically reducing the computational overhead that would otherwise be required by full Markov Chain Monte Carlo (MCMC) posterior sampling~\cite{andrieu_introduction_2003}. Finally, we employ an adaptive sampling strategy that selectively targets these high-uncertainty regions, facilitating rapid model convergence. The aforementioned methodologies are elaborated in the subsequent sections.

\subsection{Wigner function parametrisation for cat states} \label{subsec:WignerModel}
The Wigner function of an ideal cat state formed with a qcMAP protocol from a thermal state of purity $\mathcal{P}$ is \cite{yang_hot_cats_2024}
\begin{align}
    W_\mathrm{qcMAP}(\beta) = \frac{1}{\pi}\left[\mathcal{P}\left(e^{-2\mathcal{P}|\beta-\alpha|^2} + e^{-2\mathcal{P}|\beta+\alpha|^2}\right) - 2\cos\left(4\Im\{\alpha^*\beta\} +\varphi\right) e^{-2|\beta|^2/\mathcal{P}}\right].
    \label{eq:qcmap_thermal_cat_wigner}
\end{align}
An experimentally measured Wigner function of the $\alpha = 3$ cat state is shown in Fig.~\ref{fig:residuals}\textbf{(a)}. Typically, the cavity is assumed to be in the ground state, so that $\mathcal{P} = 1$. Here, we do not make this assumption and parametrise the Wigner function from Eq.~\eqref{eq:qcmap_thermal_cat_wigner} with three Gaussians
\begin{align} 
W_\mathrm{model}(\beta;\,\boldsymbol{\eta}) = A_L \mathcal{N}(\beta; \mu_L, \Sigma_L) - 2A_C \mathcal{N}(\beta; \mu_C, \Sigma_C) \cos(4\Im\{\alpha \cdot R(\phi)\beta\} + \varphi) + A_R \mathcal{N}(\beta; \mu_R, \Sigma_R),
\label{eq:model}
\end{align}
where $\mu_L, \mu_C, \mu_R \in \mathbb{R}^2$, $\Sigma_L, \Sigma_C, \Sigma_R \in \mathbb{R}^{2 \times 2}$ are the means and covariance matrices of the left, center and right Gaussians $\mathcal{N}$ with their respective amplitudes $A_L$, $A_C$, and $A_R$. The rotation matrix $R(\phi) \in \mathbb{R}^{2 \times 2}$  accounts for a correct fringe angle $\phi$ with respect to the left and right Gaussians, and $\varphi$ is the fringe phase.

\subsection{Bayesian inference}

The described model depends on the set of parameters $\boldsymbol{\eta}$ (see Appendix~\ref{app:ModelParameters}) that comprises means, variances, amplitudes, and relative angles, and our goal is to estimate those parameters from data points $\{\beta^{(i)}, W_\text{meas}(\beta^{(i)})\}_{i=1}^n$, where $\beta^{(i)}$ are complex coordinates in phase space and $W_\text{meas}(\beta^{(i)})$ are measured Wigner function values at those points. As a single-shot parity measurement is not possible in our setup, we assume that the averaged Wigner measurements are affected by normally distributed noise with variance $\sigma^2$. A standard way to estimate the model's parameters $\boldsymbol{\eta}$ in this scenario is by maximum likelihood~\cite{Bishop}, i.e.,  
\begin{align}
    \boldsymbol{\eta}_{\text{ML}} &= \argmax_\eta \mathcal{L}(\boldsymbol{\eta}), \ \text{where} \\
     \mathcal{L}(\boldsymbol{\eta}) =& \sum_{i=1}^N \log\mathcal{N}(W_\mathrm{meas}(\beta^{(i)}); W_\mathrm{model}(\beta^{(i)};\,\boldsymbol{\eta}), \ \sigma^2)
     \label{eq:sample_likelihood_detailed}
\end{align}
is the log-likelihood of the samples according to the normal distribution $\mathcal{N}$. We capture the uncertainty of $n$ measurements $\{W_\mathrm{meas}(\beta^{(i)})\}_{i=1}^n$ by assuming they are normally distributed with the mean corresponding to the model prediction $W_\mathrm{model}(\beta^{(i)};\,\boldsymbol{\eta})$ and standard deviation $\sigma$. 

Although powerful, maximum likelihood struggles in a low-data regime and tends to overfit~\cite{Bishop}. One way to compensate for this limitation is to imbue the model with prior knowledge $p(\boldsymbol{\eta})$ about the most probable model parameters, yielding a MAP objective

\begin{align}
    \boldsymbol{\eta}_{\text{MAP}} = \argmax_\eta \mathcal{L}(\boldsymbol{\eta}) + \gamma \log p(\boldsymbol{\eta}).
    \label{eq:MAP}
\end{align}
The coefficient $\gamma$ regulates the ratio between the prior and the likelihood induced by the measurements. The prior defines the Gaussians' positions as normal distributions and their variances as inverse Wishart distributions. For further details, refer to the Appendices~\ref{app:ModelParameters} and ~\ref{app:BayesianModel}.

A typical use-case setting includes defining the phase space coordinates for the measurement beforehand and using them for the state reconstruction once all the measurement results are collected \cite{ORENS, LyonNN}. However, alternating the measurement and model update steps allows to infer new phase space coordinates for the subsequent measurements, which significantly decreases the total number of required measurements and increases sensitivity to state imperfections. We introduce this idea in the following section by employing an active sampling strategy. 

\subsection{Active sampling via bootstrap} \label{subsec:Bootstrap}

Active sampling methods refine the model by interleaving parameter estimation and data collection. Crucially, data collection is performed in the regions of highest model uncertainty. Note that in our Bayesian setup, the parameters are distributed according to the posterior $p(\boldsymbol{\eta} | \text{data})$, which is not known analytically. To this end, we need to resort to an approximation. In our implementation, we use the bootstrap method \cite{Bootstrap}. Here, the uncertainty is estimated by taking the variance (i.e., \textsl{disagreement}) of $M$ models trained on $M$ different datasets $D_k, \ k \in \{1, \dots, M\}$ sampled from the original dataset $D = \{(\beta^{(1)}, W_{\mathrm{meas}}^{(1)}), \dots ,(\beta^{(n)}, W_{\mathrm{meas}}^{(n)})\}$ with replacement.

Since the models are trained on different datasets, they converge to different solutions (i.e., they disagree) and produce different Wigner maps $\{W_{\mathrm{model}}^{(j)}\}_{j=1}^M$. 

The bootstrap-generated Wigner maps are used to calculate the variance $\tilde \sigma^2$ and mean $\tilde \mu$ at each phase space point $\beta$ 
\begin{align}
    \label{eq:VarianceBootstrap}
    \tilde \sigma^2(\beta) = M^{-1} \sum_{j=1}^M \left(W_{\mathrm{model}}^{(j)}(\beta) - \tilde \mu (\beta)\right)^2, \quad \tilde \mu (\beta) = M^{-1} \sum_{j=1}^M W_{\mathrm{model}}^{(j)}(\beta;\,\boldsymbol{\eta}).
\end{align}

The variance serves as a basis for sampling subsequent points from regions of largest model uncertainty. To turn the estimated variance into a direct sampling strategy, we construct a Boltzmann distribution from the estimated standard deviation $\hat \sigma$, 
\begin{align}
    p_B(\beta) = \frac{e^{{\beta_B\tilde{\sigma}(\beta)}}}{\sum_{\beta}e^{\beta_B\tilde{\sigma}(\beta)} \text{Re}[\delta\beta] \,\text{Im}[\delta \beta]},
    \label{eq:Boltzmann}
\end{align}
where $\beta_B$ is the inverse temperature. The phase space grid is defined by $\delta\beta=\frac{\Delta I}{n_I}+i\frac{\Delta Q}{n_Q}$, where $\Delta I$ and $\Delta Q$ define the size of the sampling window and $n_I$ and $n_Q$ the number of sampling points in each direction. The Boltzmann distribution allocates more probability to higher variance regions. The inverse temperature $\beta_B$ determines the entropy of the Boltzmann distribution
\begin{align}
     H(\beta_B) \approx - \sum_{\beta} p_B(\beta) \log p_B(\beta) \text{Re}[\delta\beta] \,\text{Im}[\delta \beta],
\end{align}
with lower values of $\beta_B$ resulting in a more uniform and flatter distribution.
The new measurement points are collected by rejection sampling \cite{Bishop} from the Boltzmann distribution (Eq.~\eqref{eq:Boltzmann}). 

As training progresses, the bootstrapped models agree, yielding similar predictions and thereby reducing the variance $\hat \sigma^2(\beta)$. Consequently, the Boltzmann distribution becomes more uniform, which increases sampling from uninformative, i.e. near-zero, regions of the Wigner map (see Fig.~\ref{fig:DevMapNew}). To counteract this effect, we adjust the inverse temperature $\beta_B$ to constrain the differential entropy $H(\beta_B)$ between predetermined bounds, $H(\beta_B) \in [-2, H_\mathrm{th}]$, where $H_\mathrm{th}$ is a user-defined hyperparameter ensuring the sampling distribution does not spill too far into the uninformative regions. Similarly, the lower entropy bound prevents the distribution from becoming too sharp, thereby maintaining a reasonable acceptance rate for rejection sampling. 

\subsection{Algorithm convergence} \label{subsec:algorithm}
The reconstruction algorithm interleaves active sampling with model updating until the specified exit conditions are met (see Fig~\ref{fig:overview}\textbf{(b)}). During each iteration, the points sampled in the preceding step are measured and incorporated into the model together with all previously acquired data. These aggregated measurements are then used to construct bootstrapped datasets, which are provided as input to a model ensemble. Through MAP estimation, the ensemble generates a diverse collection of Wigner maps that serve as the basis for active sampling. The newly selected points are subsequently propagated to the next iteration of the algorithm, thereby closing the update loop. To improve computational efficiency, the actual implementation deviates slightly from the schematic representation in Fig.~\ref{fig:overview}\textbf{(b)}. Details of these modifications are provided in Appendix~\ref{app:Pipeline}.

\begin{figure}[t]
    \centering
    \includegraphics[]{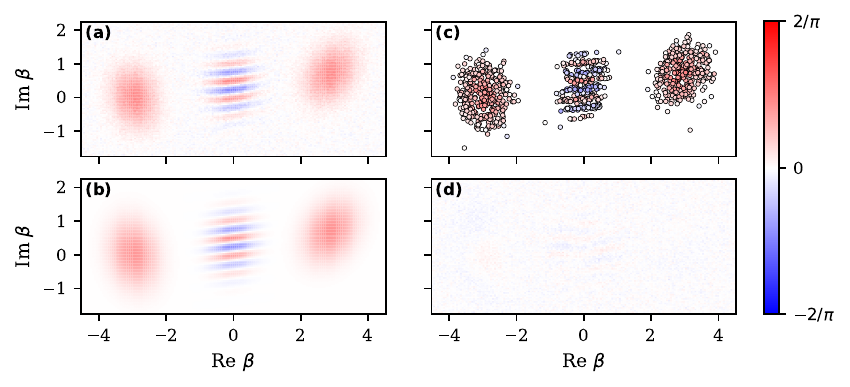}
    \caption{Example of the state reconstruction. \textbf{(a)} Reference high-resolution Wigner map. The measurement time is about 4 hours. \textbf{(b)} Reconstructed Wigner map. The measurement time is about 3.5 minutes. \textbf{(c)} 1000 measured points during the reconstruction. \textbf{(d)} Residuals between the measured and reconstructed data $R(\beta) = W_{\mathrm{meas}}(\beta) - W_{\mathrm{model}}(\beta)$. Since the model does not account for the bending of the fringes (see Eq.~\eqref{eq:model}), a small deviation is always present between the model and the observed data in the fringe area. }
    \label{fig:residuals}
\end{figure}

An example of a cat state reconstruction with size $\alpha = 3$ is shown in Fig.~\ref{fig:residuals}. Within a few minutes, the algorithm produces a Wigner map (Fig.~\ref{fig:residuals}\textbf{(b)}) that resembles all the distinct features of the high-resolution Wigner map, which conventionally requires several hours to obtain (Fig.~\ref{fig:residuals}\textbf{(a)}). The residuals between the measured and reconstructed data (Fig.~\ref{fig:residuals}\textbf{(d)}) reveal a minor discrepancy in the fringe region arising from the model parametrisation’s inability to account for fringe curvature.

\begin{figure}
    \centering
    \includegraphics[]{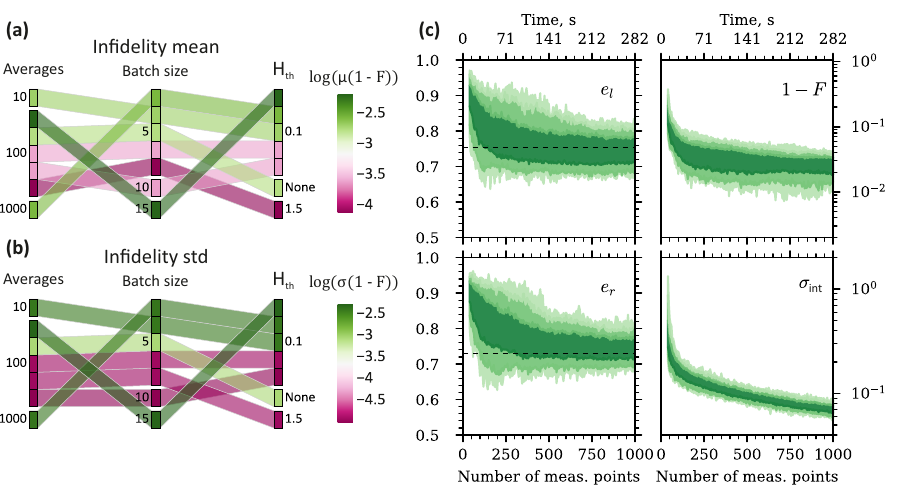}
    \caption{Hyperparameter search for the optimal reconstruction. The statistics for each set of hyperparameters is obtained from 100 experimental reconstruction runs. Infidelity is calculated as $F_2$-fidelity between the model fit to the reference high-resolution Wigner map and the reconstructed Wigner map. \textbf{(a)} Mean of infidelities, calculated by 10 bootstrapped models, after 1000 measurements for different sets of hyperparameters. Each band represents one combination of hyperparameters. Its colour shows the resulting mean infidelity, where darker purple colours represent a smaller mean infidelity across the reconstruction runs. \textbf{(b)} Standard deviation of infidelities, calculated by 10 bootstrapped models, after 1000 measurements for different sets of hyperparameters. Each band represents one combination of hyperparameters. Its colour shows the resulting standard deviation of infidelity on a logarithmic scale, where darker purple colours represent a smaller standard deviation of infidelities across the reconstruction runs. \textbf{(c)} Example of convergence curves for a hyperparameter set with 100 averages, batch size 5, and entropy threshold of 0.1. The $x$-axis corresponds to the number of points in phase space, measured during the reconstruction. The plots show ellipticities for the left and right Gaussians $e_l$ and $e_r$, infidelity, and integrated standard deviation $\sigma_\mathrm{int}$ between bootstrapped models. The shaded regions represent deviations of one, two, and three $\sigma$ going from darkest to lightest green. The dashed black lines represent the target estimand value obtained from a fit to a high-resolution Wigner map.}
    \label{fig:hyperparameters_convergence}
\end{figure}

The convergence of the reconstruction algorithm is predominantly governed by three hyperparameters: the number of averages per measurement point, the batch size, defined as the number of sampled (measured) points per iteration, and the entropy threshold used for adjusting the inverse temperature of the Boltzmann distribution. The influence of these hyperparameters on the convergence of the reconstruction algorithm is shown in Fig.~\ref{fig:hyperparameters_convergence}. We quantify the reconstruction quality by calculating the fidelity $F_\mathrm{appr}^\mathrm{(r, \ f)}$ and normalising it as  $F_2$-fidelity  \cite{Liang_2019} between the reconstructed Wigner map (r) and the model fitted to the measured high-resolution Wigner map (f), defined as:
\begin{align}
    \nonumber
    F_\mathrm{appr}^{(1, \ 2)} &= \pi \sum_{i} W^{(1)}(\beta_i) W^{(2)}(\beta_i) \text{Re}[\delta\beta] \,\text{Im}[\delta \beta],\\
    F_2 \equiv F_{2}^{\textrm{(r, f)}}&= \frac{F_\mathrm{appr}^\mathrm{(r, \ f)}}{\max \left( F_\mathrm{appr}^\mathrm{(r, \ r)}, F_\mathrm{appr}^\mathrm{(f, \ f)} \right )},  
   \label{eq:F2Reconstr}
\end{align}
where $F_\mathrm{appr}^{(1, \ 2)}$ is approximated on the phase space grid. For all the plots in Fig.~\ref{fig:hyperparameters_convergence}, the infidelity $1 - F_2$ represents the mean value predicted by $M = 10$ bootstrapped models. Each hyperparameter setting is evaluated over 100 repeated independent experimental trials to extract statistics. From here on, all fidelities in the paper are calculated using Eq.~\eqref{eq:F2Reconstr}.

Figure~\ref{fig:hyperparameters_convergence}\textbf{(a)} shows that the optimal hyperparameter configuration, yielding the lowest infidelity, consists of 100 averages, a batch size of 5, and an entropy threshold of 1.5. Among these hyperparameters, the batch size $b$ has the strongest impact on performance: increasing it to 15 significantly degrades reconstruction quality, resulting in the highest observed infidelity. This trend is expected, as larger batches provide fewer adaptive sampling steps for a fixed number of measurements, reducing reconstruction efficiency. The second most influential parameter is the entropy threshold $H_\mathrm{th}$. 
Notably, our method with entropy threshold $H_\mathrm{th}$ outperforms sampling from three Gaussians describing the cat state Wigner function (\texttt{None} in Fig.~\ref{fig:rejction_vs_no_rejection}, Appendix~\ref{app:ReconstructorConvergenceCurves}). 

For future closed-loop applications, reproducibility of the reconstruction across repeated runs is a critical requirement to ensure stable feedback and reliable control decisions. We quantify the reproducibility by evaluating the $3\sigma-$spread over 100 runs, as illustrated in Fig.~\ref{fig:hyperparameters_convergence}\textbf{(c)}. The infidelity and ellipticity plots indicate that 99.7\% (i.e., within $3\sigma$) of experimental runs yield high-fidelity reconstruction. However, this robustness does not hold for all hyperparameter configurations. For instance, with a batch size of 15, outlier runs occur in which the reconstructed maps exhibit substantially higher infidelity. Such outliers are revealed in the $3\sigma-$deviation plots, whereas considering only $2\sigma$ (95\% of runs) or $1\sigma$ (68\% of runs) spreads fails to capture them. Additional plots analogous to Fig.~\ref{fig:hyperparameters_convergence}\textbf{(c)}, showing the convergence for various hyperparameter configurations, are provided in Appendix~\ref{app:ReconstructorConvergenceCurves}.

\subsection{Early termination criteria} \label{subsec:termination}
The infidelity plot in Fig.~\ref{fig:hyperparameters_convergence}\textbf{(c)} demonstrates that different reconstruction runs require varying numbers of measurements to reach the same reconstruction fidelity. Consequently, some runs could be terminated earlier by monitoring an auxiliary metric that reflects the model’s confidence in its predictions. One such metric is the integrated standard deviation $\sigma_\mathrm{int}$ across $M$ bootstrapped models
\begin{align}
    \sigma_\mathrm{int} &= \int \tilde\sigma(\beta) d^2\beta \approx \sum_{i} \tilde\sigma(\beta_i) \text{Re}[\delta\beta] \,\text{Im}[\delta \beta]{\color{blue}.} 
\end{align}

Importantly, the convergence of $\sigma_\mathrm{int}$ behaves similarly to those of infidelity and ellipticities, as shown in Fig.~\ref{fig:hyperparameters_convergence}\textbf{(c)}. Once $\sigma_\mathrm{int}$ falls below a prescribed threshold, the reconstruction process is terminated. The $\sigma_\mathrm{int}$ threshold is set by balancing reconstruction fidelity against time constraints, as inferred from the ellipticity and infidelity plots for a given hyperparameter configuration.

\subsection{Generalisation to other states and tomography techniques}

In practical scenarios, in addition to variations in the ellipticities of the Gaussians, a cat state may undergo substantial displacements or rotations in phase space. To assess the robustness of the algorithm under such conditions, we reconstruct displaced and rotated $\alpha = 3$ cat states (see Appendix~\ref{app:ReconstructorConvergenceCurves}), using the same prior as for the non-rotated $\alpha = 3$ state benchmarked in Sec.~\ref {subsec:algorithm}. Despite the significant mismatch between the prior and the target state, the algorithm yields reliable reconstructions with high fidelity.

To further examine the generalisation capability of the reconstruction algorithm, we present convergence curves for cat states of varying sizes, $\alpha \in \{1, 1.5, 2, 2.5\}$, in Appendix~\ref{app:MoreReconstruction}. The algorithm demonstrates consistent convergence behaviour across the tested state sizes, indicating robustness with respect to $\alpha$. However, reconstructions of smaller cat states exhibit a greater dependence on stochasticity in the sampling process, requiring higher entropy thresholds to achieve comparable reconstruction fidelity.

Although our benchmarks focused on cat states, the reconstruction method is not restricted to this particular family of bosonic states. The approach could be extended to other bosonic state families such as GKP states~\cite{shi_fault-tolerant_2019}, rotation-symmetric codes~\cite{grimsmo_quantum_2020}, or general superpositions between arbitrary squeezed states~\cite{saner_generating_2026} by replacing the model parametrisation in Eq.~\eqref{eq:model} with a suitable alternative. Furthermore, the model need not be represented by a Wigner function: for example, in non-number-resolved systems, it may instead be described by a characteristic function~\cite{FastUniversalControl}. Another alternative widely used in the community, generalised Husimi $Q$ functions $Q_m$, may be less practical in this framework, as it would require additional measurements and parallel MAP estimations across different excitation numbers $m$ \cite{Kirchmair_2013, ORENS} to enhance its sensitivity to quantum features.

\section{Comparison of Wigner function sampling methods} \label{sec:SamlingComparison}

The adaptive sampling strategy yields reproducible, high-quality reconstructions suitable for closed-loop applications. Before applying it to cat state optimisation, it is instructive to place it in the broader context of existing Wigner function sampling approaches and assess their performance across different regimes.

Most Wigner function reconstruction techniques fall into two broad classes. The most widely used is grid sampling, where a phase space region of size $\Delta I \times \Delta Q$ is discretised into an $n_I \times n_Q$ grid:
\begin{equation}
    p^\textrm{gr}(\beta) = 
    \begin{cases}
    \frac{n_I n_Q}{\Delta I \Delta Q}, & \text{if} \ \textrm{Re}\beta = k\frac{\Delta I}{n_I}, \ \textrm{Im}\beta  = l\frac{\Delta Q}{n_Q}, \ k,l \in \mathbb{N}, \\
    0, &  \text{otherwise}.
    \end{cases} 
\label{eq:sampling_grid}
\end{equation}

The Wigner function is measured at each grid point, and the fidelity is approximated as $F_2 \approx F^{(\textrm{meas, target})}_\textrm{appr}$, as defined in Eq.~\eqref{eq:F2Reconstr}.

While fine grid measurements are straightforward to implement and yield reliable fidelity estimates, their cost scales unfavourably with the size of the bosonic state, leading to prohibitively long acquisition times. This overhead can be substantially reduced by employing optimal sampling strategies. Ref.~\cite{sivak_model-free_2022} derives optimal phase space sampling distributions for two distinct regimes. In the limit of a large number of measurements ($N_\mathrm{meas} \gg 1$) and for states close to the target, the optimal sampling points are drawn from the distribution
\begin{align}
    p^\textrm{sq}(\beta) \propto W_\textrm{target}^2(\beta).
    \label{eq:sampling_sq}
\end{align}
In the case where single-shot parity measurements are possible ($N_\textrm{meas} = 1$), the optimal sampling distribution becomes
\begin{align}
    p^\textrm{abs}(\beta) \propto |W_\textrm{target}(\beta)|.
    \label{eq:sampling_abs}
\end{align}
Notably, in this regime, the characterised state need not be close to the target state. This sampling strategy was successfully implemented in Ref.~\cite{LyonNN} to characterise cat states generated by a neural network. For both optimal sampling strategies defined in Eqs.~\eqref{eq:sampling_sq} and \eqref{eq:sampling_abs}, the resulting fidelity estimator takes the form
\begin{align}
    F=\pi \int W_\textrm{meas}(\beta)W_\textrm{target}(\beta) d^2\beta = \pi \mathbb{E}_{p(\beta)} \Big[ \frac{W_\textrm{target}(\beta)}{p(\beta)} W_\textrm{meas}(\beta) \Big],
    \label{eq:fidelity}
\end{align}
where $p(\beta)$ is the distribution according to which the Wigner function is sampled. As most experimental platforms allow single-shot, unbiased parity measurements, we complement the experimental results in this Section with a theoretical analysis of the fidelity uncertainty lower bound based on the Cram\'er--Rao bound (see Appendix~\ref{app:fisher-information}). The analysis shows that the optimal and adaptive sampling strategies achieve approximately 2 times lower fidelity uncertainty than grid sampling. With the experimental setup used in this paper, it is not possible to perform an unbiased single-shot measurement of the parity. Therefore, single-shot Wigner function values for $p^\textrm{abs}$ are sampled based on measured, averaged Wigner function values, as explained in Appendix~\ref{app:AbsSampling}.

We aim to estimate the fidelity within a \textsl{family} of cat states invariant under phase space rotations \(R({\phi})\) and displacements \(D(\alpha)\), rather than with respect to a single fixed cat state. There are two main motivations for this choice. First, these transformations do not qualitatively alter bosonic states: the states are identical up to a redefinition of the phase space axes and therefore should yield the same fidelity value. Second, our ultimate goal is to use the fidelity estimate in a closed-loop optimisation protocol. During such optimisation, the Wigner function of the evolving state may undergo substantial deformations, including squeezing, shifts, and rotations in phase space. 

To understand how suitable the different sampling strategies are for finding the fidelity with respect to the family of states, we consider two representative target states as a benchmark. The first is the cat state $\lvert C_{\sqrt{2}}^{\pi}\rangle$. In this case, the experimentally generated state is close to an ideal cat state, and this state is therefore used to benchmark all sampling methods defined in ~\crefrange{eq:sampling_grid}{eq:sampling_abs}, as well as the adaptive sampling strategy. The second target state is obtained by subjecting the same cat state to a phase space rotation of approximately $15^{\circ}$ and a displacement of magnitude $1.5$, i.e., $R(15^{\circ})D(1.5)\lvert C_{\sqrt{2}}^{\pi}\rangle$. This state is not close to the target state, and consequently, the sampling strategy defined in Eq.~\eqref{eq:sampling_sq} cannot be applied.

 In this section, the mean and variance of fidelity for the grid and optimal sampling strategies are estimated using a bootstrap method due to the high measurement cost. For the adaptive sampling strategy, these quantities are obtained from 100 independent experimental reconstruction runs.
  \begin{figure}[t]
     \centering
     \includegraphics[]{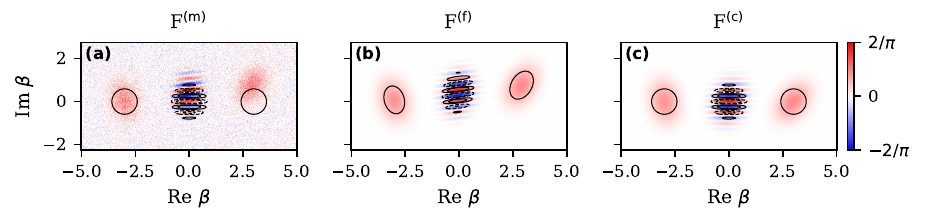}
     \caption{Illustration of the fidelity metrics (estimands) used in this section, computed according to Eq.~\eqref{eq:F2Reconstr}. The coloured Wigner map corresponds either to the measured Wigner values or to the fitted/corrected Wigner map inferred from those measurements. The black contour indicates the reference Wigner map against which the fidelity is evaluated. \textbf{(a)} Fidelity between the Wigner map of the measured state and the target cat state, $F^{(\mathrm{m})} \equiv F_{2}^{\textrm{(meas, target)}}$. \textbf{(b)} Fidelity between the fit to the measured Wigner values and the fit to a high-resolution Wigner map, $F^{(\mathrm{f})} \equiv F_{2}^{\textrm{(fit meas, fit high res)}}$. \textbf{(c)} Fidelity between the fit corrected for the transformations (displacements and rotations) and the target cat state, $F^{(\mathrm{c})} \equiv F_{2}^{\textrm{(corr, target)}}$. }
     \label{fig:fidelities_visualization}
 \end{figure}

\subsection{\texorpdfstring{$|C_{\sqrt{2}}^{\pi}\rangle$ state}{C(1.42) state}}

\begin{figure}[h]
    \centering
    \includegraphics[]{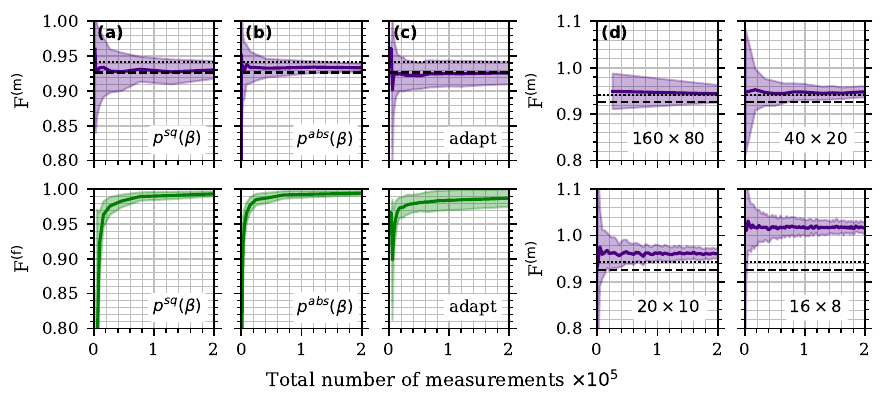}
    \caption{Comparison of Wigner function sampling strategies for $|C_{\sqrt{2}}^{\pi}\rangle$ state. Dotted line: fidelity between the measured high-resolution Wigner map and the target cat state $F = 0.94$. Dashed line: fidelity between the fit to the measured high-resolution Wigner map and the target cat state $F = 0.93$. The shaded region corresponds to $1\sigma-$deviation. Column \textbf{(a)}: sampling using the distribution $p^\textrm{sq}(\beta)$. Column \textbf{(b)}: sampling using the distribution $p^\textrm{abs}(\beta)$. Column $\textbf{(c)}$: adaptive sampling. Panel \textbf{(d)}: measurements on the grid. The grid size is fixed for each plot and specified in the plot legend. Fidelities above one are the result of insufficiently resolved Wigner maps, resulting in unphysical states.}
    \label{fig:sampling_comparison_ideal}
\end{figure}

We compare the fidelity estimates (calculated using Eq.~\eqref{eq:F2Reconstr}) obtained using different sampling methods with two reference fidelities: the fidelity between the Wigner map of the measured state and the target cat state, denoted $F^{(\mathrm{m})} \equiv F_{2}^{\textrm{(meas, target)}}$, and the fidelity between a fit to the Wigner values measured up to now and the fit to a high resolution Wigner map of the cat state, denoted $F^{(\mathrm{f})} \equiv F_{2}^{\textrm{(fit meas, fit high res)}}$ (see illustration in Fig.~\ref{fig:fidelities_visualization}). The results shown in Fig.~\ref{fig:sampling_comparison_ideal} demonstrate that both the optimal sampling strategy $p^{\mathrm{sq}}(\beta)$ and the adaptive sampling strategy converge to the same target fidelity value. The optimal sampling strategy $p^{\mathrm{abs}}(\beta)$ converges to a slightly different value, even though it is expected to yield an unbiased estimate. We attribute this discrepancy to drifts in the calibration of the Wigner function measurement scaling over the extended measurement time required by this sampling strategy (see Appendix~\ref{app:AbsSampling} for details). Finally, $F^{(\mathrm{f})}$ approaches unity for all three methods, with the adaptive sampling strategy exhibiting a slightly larger standard deviation. We note that the larger observed standard deviation can be partially attributed to the method used to estimate the fidelity uncertainty. The adaptive sampling results are obtained from 100 independent experimental reconstructions, whereas the optimal sampling results rely on bootstrapped data, which may underestimate the true uncertainty. This interpretation is consistent with the Cram\'er--Rao analysis presented in Appendix~\ref{app:fisher-information}, which predicts that, for $\alpha=3$ cat states and sufficiently large measurement budgets, the uncertainty lower bound of the adaptive sampling strategy approaches that of the $p^\textrm{abs}$ sampling strategy.

We characterise the performance of the grid sampling as follows. Starting from the high-resolution Wigner map used to compute the reference fidelities, we downsample the map and vary the total number of measurements by increasing the number of averages per grid pixel (see Appendix~\ref{app:GridSampling} for corresponding Wigner maps). The results shown in Fig.~\ref{fig:sampling_comparison_ideal} demonstrate that coarser grids lead to biased fidelity estimates and, in extreme cases, give unphysical estimates.

\subsection{\texorpdfstring{$R(15^{\circ})D(1.5)\lvert C_{\sqrt{2}}^{\pi}\rangle$ state}{R(15)D(1.5)C(1.41)} state}

We perform a similar analysis for a cat state of the same size and phase as in the previous section, rotated by $15^{\circ}$ and displaced by $1.5$ in phase space. In addition to the fidelities $F^{(\mathrm{m})}$ and $F^{(\mathrm{f})}$ introduced in the previous section, we also consider $F^{(\mathrm{c})} \equiv F_{2}^{\textrm{(corr, target)}}$, the fidelity between the fit corrected for the transformations and the target cat state. Owing to invariance under these transformations, $F^{(\mathrm{c})}$ is expected to converge to the same target fidelity values stated in Fig.~\ref{fig:sampling_comparison_ideal} in the absence of measurement drifts.

\begin{figure}
    \centering
    \includegraphics[]{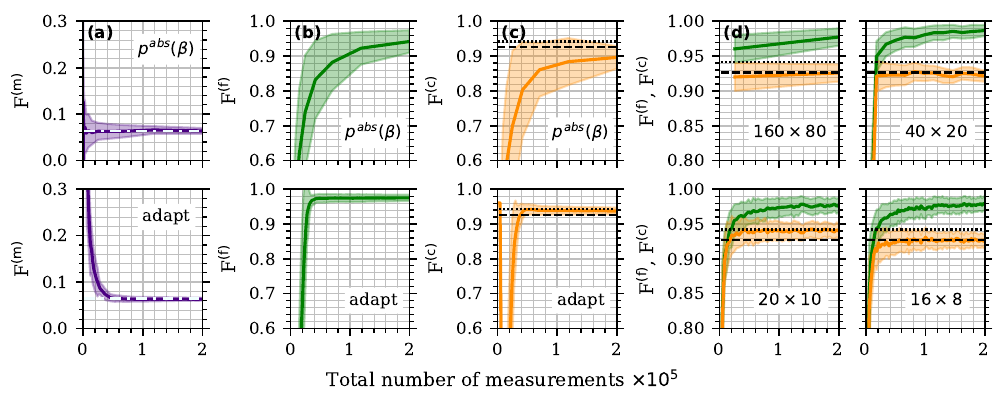}
    \caption{Comparison of Wigner function sampling strategies for a rotated and displaced state $R(15^{\circ})D(1.5)\lvert C_{\sqrt{2}}^{\pi}\rangle$. Column \textbf{(a)}: fidelity $F^\mathrm{(m)}$ between the measured Wigner map and the target cat state $\lvert C_{\sqrt{2}}^{\pi}\rangle$ for adaptive and optimal sampling $p^\textrm{abs}(\beta)$. Dash-dot line: fidelity between the measured high-resolution Wigner map and the target cat state $\lvert C_{\sqrt{2}}^{\pi}\rangle$. Column $\textbf{(b)}$: fidelity $F^\mathrm{(f)}$ between the fit to the measured Wigner values (adaptive or optimal sampling $p^\textrm{abs}(\beta)$) and the fit to a high-resolution Wigner map. Column \textbf{(c)}: fidelity $F^\mathrm{(c)}$ between the fit corrected for the transformations and the target cat state $\lvert C_{\sqrt{2}}^{\pi}\rangle$ for adaptive and optimal sampling $p^\textrm{abs}(\beta)$. Dotted and dashed lines as described in Fig.~\ref{fig:sampling_comparison_ideal}. Panel \textbf{(d)}: $F^\mathrm{(f)}, \ F^\mathrm{(c)}$ for the grid sampling. The grid size is fixed for each plot and specified in the plot legend. Fidelity $F^\mathrm{(f)}$ is plotted in green, $F^\mathrm{(c)}$ is plotted in orange. The shaded region corresponds to $1\sigma-$deviation. }
    \label{fig:sampling_comparison_displ}
\end{figure}

Fig.~\ref{fig:sampling_comparison_displ} shows that, for this state, the adaptive sampling strategy achieves significantly faster convergence of $F^{(\mathrm{f})}$ and successfully predicts $F^{(\mathrm{c})}$, whereas the optimal sampling strategy fails to reconstruct the state accurately. This failure is primarily due to the absence of sampled points in the region of phase space occupied by the displaced cat state (see Fig.~\ref{fig:sampling_abs_steps} in Appendix \ref{app:AbsSampling} for corresponding Wigner maps).

The grid sampling area is chosen so that it contains the Wigner function features of the cat state (see Fig.~\ref{fig:grid_undersampling_displ} in Appendix~\ref{app:GridSampling} for corresponding Wigner maps). In this case, both $F^{(\mathrm{f})}$ and $F^{(\mathrm{c})}$ are close to the values calculated from the high-resolution Wigner map. Naturally, the estimates are correct only if a relevant region of phase space is chosen.

\subsection{Sampling strategies overview}

Grid sampling remains a standard tool for Wigner tomography of bosonic states and provides unbiased fidelity estimates when the relevant phase space region and grid resolution are known in advance. This makes it well-suited for stand-alone state characterisation with weak time constraints. However, in closed-loop optimisation, the available reconstruction time is typically limited to a few minutes, while the state may undergo substantial displacements in phase space, which renders grid sampling impractical.

Sampling from the absolute value of the Wigner function, $p^\text{abs}(\beta)$, enables direct estimation of $F^{\mathrm{(m)}}$ but loses efficiency as the state is displaced away from the optimal sampling region. As a consequence, the accuracy of the $F^{\mathrm{(f)}}$ estimate degrades, which makes $p^\text{abs}(\beta)$ unsuitable for closed-loop optimisation. Furthermore, single-shot parity measurements are not universally available across experimental platforms. When parity is inferred from averaged measurements at each phase space point, sampling according to $p^\text{sq}(\beta)$ is required, a strategy that is valid only for states close to ideal ones. $p^\text{sq}(\beta)$ cannot be applied to rotated or displaced states in the phase space, making it unsuitable for closed-loop applications as well.

By contrast, the introduced adaptive reconstruction approach combines adaptive sampling with iterative model updates and therefore performs robustly across both regimes considered here. It enables efficient extraction of fidelity with respect to a family of bosonic states, even in the presence of large phase space displacements. In addition, as discussed in Sec.~\ref{sec:Reconstruction}, the method is inherently sensitive to subtle deviations of the reconstructed state, such as Gaussian ellipticity, which are not captured by fidelity alone. We emphasise that we did not perform a hyperparameter search for the adaptive reconstruction in this section, and the algorithm's performance may still be optimised.

Alternative approaches to bosonic state characterisation include ORENS~\cite{ORENS}, which reconstructs the density matrix from generalised Husimi $Q$ measurements~\cite{Kirchmair_2013}. However, the original work does not report the run-to-run variance of the reconstructed fidelity, and the experimental demonstrations are limited to small bosonic states, well confined within a Hilbert space of dimension $D = 6$. For our example state $R(15^{\circ})D(1.5)\lvert C_{\sqrt{2}}^{\pi}\rangle$, about $50\%$ of the population is contained within that limited Fock space, requiring a dimension of $D=12$ to cover $90\%$ of the population. Hence, a direct comparison between the methods is difficult. Similar to optimal sampling strategies, ORENS is well-suited for estimating the fidelity of bosonic states centred in phase space, but it is not designed to evaluate fidelity across a family of bosonic states related by phase space transformations and not expected to be sensitive to subtle state deviations.

A conceptually different route to bosonic state tomography is provided by generative adversarial networks~(GANs). In Ref.~\cite{PhysRevLett.127.140502}, the authors report extensive benchmarking on simulated cat states with $\alpha \in [1,3]$, achieving infidelities comparable to those obtained in the present work for similar numbers of measurement points. The tomography points (i.e., phase space displacements) are selected randomly prior to reconstruction rather than adaptively during the measurement process. Employing an active learning approach similar to the one presented in this paper would improve the GAN-based approach.

\section{Closed-loop QOC using adaptive reconstruction} \label{sec:ClosedLoop}
As a proof-of-principle application, we employ the reconstruction technique to evaluate a FoM in a closed-loop QOC setting. In this approach, the control pulses are iteratively refined using feedback from experimental data (Fig.~\ref{fig:overview}\textbf{(a)}) to improve the fidelity of cat states of varying sizes and parities. In contrast to open-loop approaches, which depend on accurate knowledge of the system Hamiltonian, closed-loop strategies are robust to system drift, measurement noise, and modelling inaccuracies. This robustness comes at the expense of efficiency, as closed-loop methods generally require a larger number of measurement points to gather sufficient information about the state of the bosonic system.

At every iteration of the closed loop, the optimiser modifies the generation protocol by replacing the waiting time with qubit control pulses and tuning the phase of the first qubit pulse (Fig.~\ref{fig:overview}\textbf{(c)}). As an initial guess, we use a zero-amplitude qubit pulse and set the phase of the first pulse to $\phi = 0$. These pulse settings are passed to the reconstructor (Fig.~\ref{fig:overview}\textbf{(b)}), which executes the experiment until a reliable model of the state is obtained. The FoM is subsequently evaluated (Fig.~\ref{fig:overview}\textbf{(a)}) as the $F_2$ fidelity between the reconstructed and target states (see Eq.~\eqref{eq:F2Reconstr}), augmented with additional penalty terms (see Sec.~\ref{sec:fom}). This FoM is returned to the optimiser, which generates the next candidate pulse sequence.

We used the dCRAB (dressed Chopped Random Basis) method~\cite{Caneva2011,Rach2015} in combination with the Nelder–Mead updating algorithm from the QuOCS package~\cite{rossignolo} as the closed-loop optimiser. The dCRAB approach defines the pulse parametrisation and has been widely applied across diverse systems~\cite{muller_one_2022}, whereas the updating algorithm governs how the optimisation parameters are iteratively adjusted~\cite{nelder_simplex_1965}.

In the dCRAB formalism, each control pulse is parametrised with a set of random basis functions $f(\alpha_j,t), \ j \in [1, N_c]$ defined by superparameters $\alpha_j$. These superparameters are initialised at the start of the optimisation and can be restricted to a prescribed interval, $\alpha_j\in [\alpha_\text{min},\,\alpha_\text{max}]$, thereby defining the family of admissible pulses. The optimisation is then performed over the weights $A_j$ associated with the basis functions, which span the parameter space explored by the updating algorithm. Before suggesting a new set of parameters, the updating algorithm is given a FoM associated with the control pulse. To balance the number of measurements and the noise, anytime the FoM comes close to the optimum, it is remeasured to confirm the outcome with respect to the previous best result. The control pulse $u(t)$ is constructed by adding the weighted sum of basis functions to the initial guess $u_0(t)$, yielding
\begin{equation}
    u(t) = u_0(t) + \sum_{j=1}^{N_c} A_j f(\alpha_j;t).
    \label{eq:iterations}
\end{equation}
To enable further improvements without increasing the dimensionality of the parameter space, dCRAB employs super-iterations. Each super-iteration comprises a full optimisation cycle, beginning with the previously optimised pulse, 
$u_0(t) = u_\mathrm{opt}(t)$, which is combined with a newly randomised set of basis functions $\{f(\alpha_j; t)\}_{j = 1}^{N_c}$. The parameters $A_j$ in Eq.~\eqref{eq:iterations} are then optimised according to the chosen updating algorithm. At the start of each super-iteration, the previously obtained optimum is remeasured to compensate for slow system drifts and to reduce the impact of statistical outliers.

Prior to performing the closed-loop optimisation, we benchmarked updating algorithms, basis sets, and super-parameter ranges using a gradient-free open-loop optimisation. The tested updating algorithms included the $1+1$ algorithm~\cite{kern_1+1_2004}, CMA-ES~\cite{hansen_cma_2023}, and Nelder–Mead~\cite{nelder_simplex_1965}, while the basis families considered were Fourier, Sinc, and Sigmoid~\cite{pagano_role_2024}. The remaining hyperparameters depend on the choice of the updating algorithm. The optimal configuration for closed-loop optimisation is summarised in Table~\ref{tab:ClosedLoopConfig}.

\subsection{Configuration of the optimiser}\label{sec:fom}

At each optimisation step, for a given set of control pulses and parameters, the reconstructor returns a parametric model of the reconstructed state (Eq.~\ref{eq:model}). The parameters used for the reconstruction are summarised in Appendix~\ref{app:OptimalParameters}. Since the displacement and rotation of the cat state in phase space can be directly extracted from the model parameters, they can be used to construct a corrected Wigner map. This correction enables the evaluation of the state fidelity with respect to the entire family of cat states, which is invariant under phase space translations and rotations. 
We calculate the fidelity $F^\mathrm{(c)}$ between the corrected reconstructed and target states as defined in Eq.~\eqref{eq:F2Reconstr}. 

As feedback to the optimiser, we construct a FoM that incorporates both the corrected state fidelity $F^\mathrm{(c)}$ and a penalty term $P$ accounting for state imperfections. This helps to augment the landscape to avoid local minima. Open-loop tests showed that a quadratic dependence of the FoM on the fidelity helps to correct Kerr-induced distortions. Accordingly, the optimiser is provided with a FoM of the form \(\text{FoM} = F^{\mathrm{(c)}^2} - P \).

To put more emphasis on the errors introduced by the cavity Kerr $K_c$, we consider various penalty terms that account for the ellipticities of the left (\textit{l}) and right (\textit{r}) Gaussians, $e = \sigma_{\mathrm{min}}/\sigma_{\mathrm{max}}$. We characterise penalty terms by calculating the signal-to-noise ratio (SNR) at the end of the reconstruction process, as summarised in Table~\ref{tab:SNR_penalties}. To smooth the optimisation landscape, we minimise FoM noisiness by selecting the penalty term associated with the highest SNR, i.e.
\begin{align}
    P = \sum_{i \in [l, r]} \left(1 - e_i^2 \right)^{1/2}.
    \label{eq:PClosedLoop}
\end{align}

\begin{table}[h!]
    \centering
    \captionsetup[subtable]{labelformat=empty} 
    \begin{subtable}[t]{0.65\linewidth}
        \centering
        \llap{\textbf{a) }\hspace{-10mm}} 
        \vspace{0pt} 
        \begin{tabular}{c|c}
            \hline
            \textbf{Optimisation parameter} & \textbf{Value} \\ 
            \hline
            basis & sinc with $\omega_\mathrm{max} \in [1, 300]$ \\
            \hline
            basis vector number & 4 \\
            \hline
            basis max & 2 \\
            \hline
            amplitude variation & $20\%$ \\
            \hline
            superiteration number & 2\\
            \hline
            maximum number of evaluations & 1000 \\
            \hline
        \end{tabular}
        \caption{} 
        \label{tab:ClosedLoopConfig}
    \end{subtable}
    \hfill
    \begin{subtable}[t]{0.32\linewidth}
        \centering
        \llap{\textbf{b) }\hspace{0mm}} 
        \vspace{0pt} 
        \begin{tabular}{ c | c | c | c } 
            \hline
             $P_i, i \in [l, r]$ & Mean & Std & SNR \\ 
            \hline 
            $1-e_i$ & 0.24 & 0.02 & 10.60 \\ 
            \hline 
            $\sqrt{1-e_i}$ & 0.49 & 0.02 & 20.74 \\ 
            \hline 
            $\sqrt{1-e_i^2}$ & 0.65 & 0.03 & 23.84 \\ 
            \hline 
            $1-e_i^2$ & 0.43 & 0.04 & 12.17 \\ 
            \hline 
            $(1-e_i^2)^2$ & 0.18 & 0.03 & 6.31 \\ 
            \hline
            $(1-e_i)^2$ & 0.06 & 0.01 & 5.49 \\ 
            \hline
        \end{tabular}
        \caption{} 
        \label{tab:SNR_penalties}
    \end{subtable}
    \captionsetup{justification=raggedright,singlelinecheck=false}
    \caption{%
        \textbf{a)} Optimal configuration for closed-loop optimisation using the Nelder-Mead algorithm and dCRAB ansatz in accordance with the QuOCS framework~\cite{rossignolo}. \textbf{b)} Mean, standard deviation, and signal-to-noise ratio (SNR) for different penalty functions $P = \sum_{i \in [l, r]} P_i$ after 1000 iterations of reconstruction. The penalty with the largest SNR (3rd row) is the optimal choice for constructing the FoM.
    }
\end{table}

\subsection{Closed-loop runs with varying target states and FoMs}

\begin{figure}[h]
    \centering
    \includegraphics[]{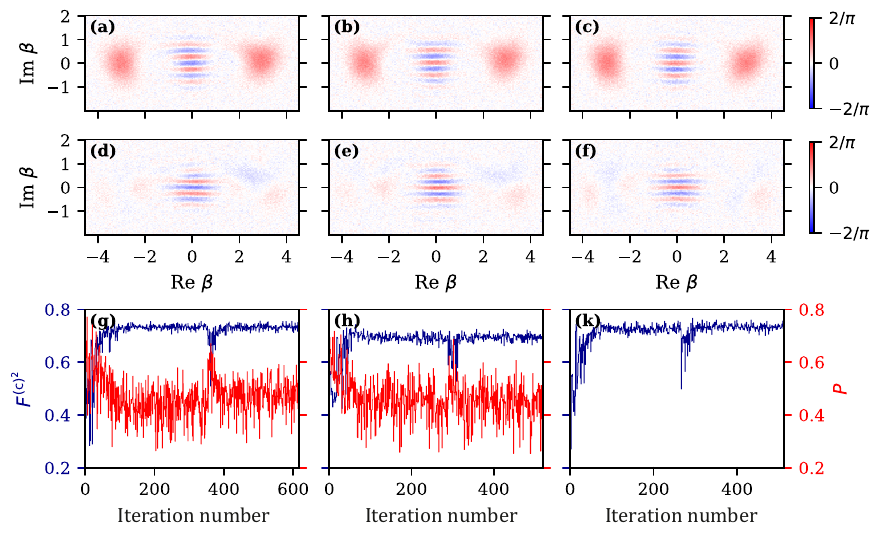}
    \caption{Closed-loop optimisation of $\alpha = 3$ cat state with zero amplitude initial guess for the correction pulse. The first row displays the Wigner maps of the optimised states, the second row shows the difference between the target cat state and the optimised state, and the third row shows $F^{\mathrm{(c)}^{2}}$ and the penalty as a function of the iteration number. Each column corresponds to one run. The runs exploit different figures of merit and target the cat states with different phases $\varphi$ (parities). The residuals observed in Fig.~\textbf{(d–f)} arise from a reduction in fringe amplitude caused by decoherence. \textbf{(a)}, \textbf{(d)}, and \textbf{(g)}: $\mathrm{FoM} = F^{\mathrm{(c)}^2} - P, \ \varphi = 0$. \textbf{(b)}, \textbf{(e)} and \textbf{(h)}: $\mathrm{FoM} = F^{\mathrm{(c)}^2} - P, \ \varphi = -\pi$. \textbf{(c)}, \textbf{(f)} and \textbf{(k)}: $\mathrm{FoM} = F^{\mathrm{(c)}^2}, \ \varphi = -\pi$. The relevant FoM parameters, final fidelity and ellipticity are summarised in Table \ref{tab:ClosedLoopSummary}.}
    \label{fig:closed_loop_runs}
\end{figure}

We evaluated the adaptability of the closed-loop optimisation across different target cat sizes $\alpha$, fringe phases $\varphi$ and FoMs. A summary of the results is provided in Table~\ref{tab:ClosedLoopSummary}. Across all runs, the optimisation consistently reduced Kerr-induced distortions, reflected in improved Gaussian ellipticities. Although the inclusion of a penalty term introduces additional noise into the FoM, it leads to a substantially greater improvement in ellipticity by the end of the optimisation. Representative cat states and convergence curves for selected runs are shown in Fig.~\ref{fig:closed_loop_runs}.

In addition to reducing Kerr distortions, most optimisation runs also improved the preparation fidelity of the cat state. For instance, comparing runs \#3 and \#4 shows that including a penalty term in the FoM leads to states with improved Gaussian ellipticity, while the unpenalised optimisation achieves slightly higher fidelity. A similar trade-off can be observed in run \#5, where Kerr distortions were mitigated at the expense of a small reduction in fidelity.

These behaviours are consistent with the characteristics of gradient-free optimisation methods. Updating algorithms such as Nelder–Mead are typically efficient in relatively low-dimensional parameter spaces but have a limited ability to explore more complex optimisation landscapes~\cite{sorensen_quantum_2018}. At the same time, the dimension of the solution scales with the problem dimension~\cite{caneva_complexity_2014, PhysRevLett.113.010502}. As a result, identifying pulse shapes that simultaneously minimise Kerr distortions and maximise fidelity can be challenging, and different optimisation settings may favour distinct aspects of the target state.

\begin{table}[h]
\makebox[\textwidth][c]{
    \begin{tabular}{c|c|c|c|c|c|c|c}
        \hline
         Run \# & $\alpha$ & $\varphi$ & FoM & $F^\mathrm{(c)}$, init & $F^\mathrm{(c)}$, final & $(e_l, e_r)$, init & $(e_l, e_r)$, final \\
         \hline
         1 &3 & 0 & $F^2 - P$ & 0.61 & 0.74 & (0.73, 0.73) & (0.79, 0.98) \\ 
         \hline
         2 & 3 & $-\pi/2$ & $F^2 - P$ & 0.29 & 0.71 & (0.73, 0.73) & (0.72, 0.98) \\ 
         \hline
         3 & 3 & $-\pi$ & $F^2 - P$ & 0.40 & 0.71 & (0.73, 0.73) & (0.79, 0.94) \\ 
         \hline
         4 & 3 & $-\pi$ & $F^2$ & 0.40 & 0.75 & (0.73, 0.73) & (0.73, 0.77) \\ 
         \hline
         5 & 2.5 & 0 & $F^2 - P$ & 0.82 & 0.79 & (0.81, 0.82) & (0.96, 0.96) \\ 
         \hline
         6 & 2.5 & $-\pi$ & $F^2 - P$ & 0.21 & 0.78 & (0.81, 0.82) & (0.91, 0.96) \\ 
         \hline
    \end{tabular}
}
\caption{Closed-loop optimisation runs for different cat sizes $\alpha$, target cat phases $\varphi$, and FoMs. All the runs for $\alpha = 3$ ($\alpha = 2.5$) start from the same cat, corrected for an offset rotation and displacement. Initial and final states are characterised with $F_2$ fidelity between the measured and target cat states (see Eq.~\ref{eq:F2Reconstr}). Right and left ellipticities are denoted as $e_l = \sigma_\mathrm{min}^l/\sigma_\mathrm{max}^{l}, \ e_r = \sigma_\mathrm{min}^r/\sigma_\mathrm{max}^{r}$}
\label{tab:ClosedLoopSummary}
\end{table}

\section{Conclusion} \label{sec:Conclusion}
We have introduced a reconstruction technique that enables the estimation of bosonic state fidelity within a family of states invariant under phase space translations and rotations on a timescale of only a few minutes. We demonstrate that the reconstruction is reproducible across trials and sensitive to subtle state imperfections by analysing statistics from 100 independent experimental runs on cat states of varying sizes, with $\alpha \in [1,3]$. We further present an extensive study of the impact of reconstruction hyperparameters on convergence, together with a comparison of existing Wigner function sampling strategies. While the current implementation already achieves short reconstruction times, further speedups are feasible, for example, by updating only displacement values rather than the full configuration of the experiment control system, employing more efficient sampling algorithms for the Boltzmann distribution in Eq.~\eqref{eq:Boltzmann}, or migrating computationally intensive parts of the code from Python to C++ or FPGA.

Although the present study focuses on reconstructing cat states from measured Wigner values, the proposed reconstruction technique is not restricted to a particular class of bosonic states or to a specific tomography method. More generally, the algorithm could be applied to arbitrary bosonic states and tomography schemes, provided that an appropriate model definition and parametrisation are chosen. We emphasise that closed-loop applications of reconstruction techniques must satisfy three key requirements identified in Sec.~\ref{sec:Reconstruction}: small estimation error, reproducibility, and short total reconstruction time on the order of a few minutes. Previously, these requirements have not been simultaneously achieved. This makes our technique a useful tool for a wide range of experiments with bosonic states. As proof of principle, we demonstrate a successful implementation of our reconstruction technique in a closed-loop optimisation setting. 

While the closed-loop results presented here are modest, they establish a solid foundation for future closed-loop experiments on bosonic states. We attribute the limited improvement of the cat states, aside from low coherence times, to the restricted ability of direct-search optimisation algorithms to efficiently explore a high-dimensional and structured control landscape. Overcoming this limitation requires more sophisticated optimisation frameworks capable of exploiting the underlying structure of the quantum dynamics.

A particularly promising direction for future work is the application of physics-informed reinforcement learning (RL)~\cite{PhysicsInformedRLReview}. One approach is model augmentation, whereby a model of the environment (in the RL sense, encompassing both the quantum system and its surrounding bath) is learned by embedding the Master or Schrödinger equation into a neural network architecture, similarly to Lagrangian neural networks~\cite{RameshLagrangianNet, Cranmer2020LagrangianNN}. Such a model could generate synthetic trajectories for training, substantially reducing the number of costly experimental evaluations.

Another avenue is policy augmentation~\cite{PhysicsInformedRLReview}. In particular, neural networks could be used to learn higher-order Hamiltonian contributions and refine pre-characterised system parameters, which would subsequently be passed to the Schrödinger equation governing the system dynamics. The resulting control policies would therefore be constrained by solutions to physically consistent differential equations that describe the underlying quantum evolution. Combined with the reconstruction framework developed here, physics-informed RL approaches could enable a new generation of adaptive control protocols for bosonic quantum systems.

\section*{Data availability}
The data that support the findings of this study are available at Zenodo 10.5281/zenodo.21372891.

\section*{Code availability}
The code for the reconstruction is available on GitHub:

https://github.com/Vasilisa-Usova/adaptive-reconstruction/

\section*{Acknowledgements}
This research was funded in part by the Austrian Science Fund (FWF) under DOI 10.55776/F71 within the BeyondC SFB and under DOI 10.55776/ESP4563925 (LeQuaC). We further acknowledge funding from the Austrian Federal Ministry of Education, Science and Research via the Austrian Research Promotion Agency (FFG) through the flagship project FO999897481 (HPQC) and FO999914030 (MUSIQ). SM acknowledges funding from the Quantum Technology Flagship project PASQuanS2.1, project grant agreement No. 101113690. For the purpose of open access, the author has applied a CC BY public copyright licence to any Author Accepted Manuscript version arising from this submission. 
\clearpage

\appendix
\appendixfigures
\appendixtables
\section*{Appendix}
\pdfbookmark[1]{Appendix Contents}{app}

\begin{enumerate}
\appsection{app:Reconstructor}{Appendix A: Reconstructor}
\appsubsection{app:ModelParameters}{A.1 Model parameters}
\appsubsection{app:BayesianModel}{A.2 Bayesian model}
\appsubsection{app:boltzmann}{A.3 Sampling from high variance regions}
\appsubsection{app:Pipeline}{A.4 Pipeline}
\appsubsection{app:OptimalParameters}{A.5 Optimal parameters}

\appsection{app:Sampling}{Appendix B: Sampling strategies}
\appsubsection{app:GridSampling}{B.1 Grid Sampling}
\appsubsection{app:AbsSampling}{B.2 $p^\mathrm{abs}$}

\appsection{app:ReconstructorConvergenceCurves}{Appendix C: Reconstruction convergence for $\alpha = 3$ cat state}

\appsection{app:MoreReconstruction}{Appendix D: Reconstruction convergence for smaller cat states}

\appsection{app:ExpSetup}{Appendix E: Experimental Setup}

\appsection{app:fisher-information}{Appendix F: Fisher Information and Cramér-Rao Bound}

\end{enumerate}

\clearpage

\clearpage
\section{Reconstructor} 
\label{app:Reconstructor}

\subsection{Model parameters} 
\label{app:ModelParameters}
\begin{figure}[h!]
    \centering
    \includegraphics{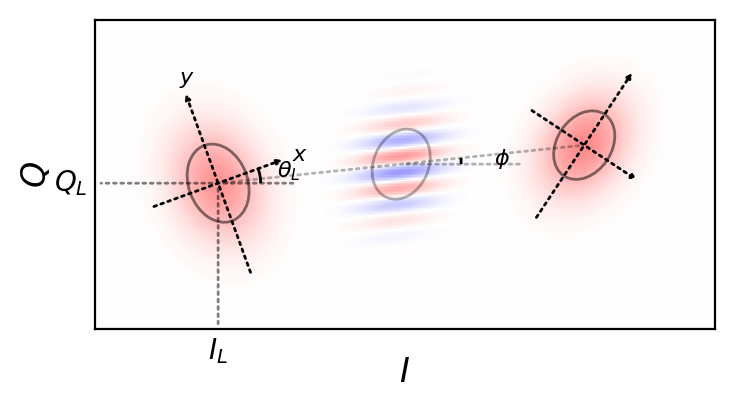}
    \caption{Illustration of the model parameters on the Wigner map. For each covariance matrix, eigenvalues and eigenvectors are calculated. Based on that, the minimum and maximum standard deviations $\sigma_x$ and $\sigma_y$, and the angle between the principal axis and the Cartesian x-axis are evaluated.}
    \label{fig:Model}
\end{figure}

As described in Sec.~\ref{subsec:WignerModel}, the Wigner function is parametrised according to Eq.~\eqref{eq:model}, resulting in 16 free optimisation parameters (see Fig.~\ref{fig:Model}),
\begin{align}
\label{eq:set_of_model_parameters}
\mathbf{\eta} = \{&I_L, Q_L, \sigma_x^{(L)}, \sigma_y^{(L)}, \theta_L, \\ \nonumber\
&I_R, Q_R, \sigma_x^{(R)}, \sigma_y^{(R)}, \theta_R, \\ \nonumber\
&\sigma_x^{(C)}, \sigma_y^{(C)}, \theta_C, \\ \nonumber\
&A_L, A_C, \varphi\}.
\end{align}
Here, $\mu_{L(R)} = (I_{L(R)}, Q_{L(R)})$ denotes the mean of the left (right) Gaussian. The parameters $\sigma_x^{(L,R,C)}$ and $\sigma_y^{(L,R,C)}$ are the corresponding standard deviations in the coordinate system defined by the principal axes, rotated by an angle $\theta_{L(R,C)}$ with respect to the Cartesian axes. The parameters $A_{L(C)}$ denote the amplitudes of the left (central) Gaussians, and $\varphi$ is the phase of the cat state. 
The remaining parameters are expressed in terms of those above; in particular, the fringe angle $\phi$ and the centre of the cat state $\mu_C$ in phase space are
\begin{align}
\phi &= \arctan \frac{Q_R - Q_L}{I_R - I_L}, \\
\mu_C &= 0.5(\mu_L + \mu_R).
\label{eq:middle_center}
\end{align}
Normalisation of the cat state density matrix, $\mathrm{tr} \rho = 1$, imposes
\begin{align}
A_R = 1 - A_L.
\end{align} 

During the maximum a posteriori (MAP) estimation (see Sec.~\ref{subsec:WignerModel}), the parameters are constrained to the following bounds:
\begin{align}
\mu_{L,R} &\in [-4, 4], \nonumber \\
\sigma_x^{(L,R,C)}, \sigma_y^{(L,R,C)} &\in [0.25, 1], \nonumber \\
\theta^{(L,R,C)} &\in \left[-\frac{\pi}{2} - \epsilon, \frac{\pi}{2} + \epsilon\right], \nonumber \\
A_L &\in [0.5 - A_\mathrm{imb} - \epsilon, 0.5 + A_\mathrm{imb} + \epsilon], \nonumber \\
A_C &\in [0.1, 0.5 + \epsilon],
\label{eq:bounds}
\end{align}
where $A_\mathrm{imb} = 0.05$ accounts for a possible amplitude imbalance between the left and right Gaussians, which may arise, for example, from imperfect qubit disentanglement, and $\epsilon = 10^{-3}$ is a small offset that allows the optimisation algorithm to reach the boundaries of the allowed parameter ranges. During the reconstruction tests, we imposed as well a constraint on the cat state phase $\varphi \in [-\pi - \epsilon, \pi + \epsilon]$. However, we later found that, for some phases, the optimisation tends to get stuck at the boundary, so we lifted this assumption to increase stability. 

\subsection{Bayesian model} 
\label{app:BayesianModel}

The MAP objective (see Eq.~\eqref{eq:MAP}) requires a prior distribution $p(\eta)$ over the model parameters $\mathbf{\eta}$, which we specify as follows. At the beginning of the reconstruction, we assume that the experimentally prepared cat state is close to an ideal one. In particular, the centres of the left and right Gaussians are assumed to lie near their ideal phase-space positions at $\pm \alpha$,
\begin{align*}
\mu_L &\sim \mathcal{N}(\mu_L^\mathrm{prior} = (-\alpha, 0), \ \Sigma_L^\mathrm{prior} = \mathrm{diag}(0.25, 0.25)), \\
\mu_R &\sim \mathcal{N}(\mu_R^\mathrm{prior} = (\alpha, 0), \ \Sigma_R^\mathrm{prior} = \mathrm{diag}(0.25, 0.25)).
\end{align*}
The covariance matrices are assigned a high-entropy inverse Wishart prior,
\begin{align*}
\mathcal{W}^{-1}_{L} = \mathcal{W}^{-1}_{R} = \mathcal{W}^{-1}_{C} = \mathcal{W}^{-1}(\psi = \mathrm{diag}(0.25, 0.25), \ \nu = 1),
\end{align*}
which suppresses unphysical covariance estimates while avoiding excessive bias toward the prior.

Collecting all contributions, the prior term in Eq.~\eqref{eq:MAP} takes the form
\begin{align}
\log p(\eta) = &\log\mathcal{N}(\mu_L; \mu_L^\mathrm{prior}, \Sigma_L^\mathrm{prior})
+ \log\mathcal{N}(\mu_R; \mu_R^\mathrm{prior}, \Sigma_R^\mathrm{prior}) + \nonumber \\
&\log\mathcal{W}^{-1}(\Sigma_L; \psi_L^\mathrm{prior}, \nu_L^\mathrm{prior})
+ \log\mathcal{W}^{-1}(\Sigma_R; \psi_R^\mathrm{prior}, \nu_R^\mathrm{prior}) + \nonumber \\
&\log\mathcal{W}^{-1}(\Sigma_C; \psi_C^\mathrm{prior}, \nu_C^\mathrm{prior}).
\label{eq:parameter_likelihood}
\end{align}
As additional measurement data are acquired during the reconstruction, the relative weight of the log-likelihood term increases, thereby reducing the influence of the prior.

In the numerical implementation, we employ a slightly modified MAP objective that accounts for the number of measurements per iteration $b$ (batch size),
\begin{align}
\mathbf{\eta}_{\mathrm{MAP}} = -\arg\min_\eta \left(\mathcal{L}(\eta)/b\right) - \gamma \log p(\eta),
\label{eq:MAP_modified}
\end{align}
 where normalisation of the likelihood by the batch size ensures comparable convergence behaviour for different values of $b$, and $\gamma$ controls the impact of the prior.

\subsection{Sampling from high variance regions} \label{app:boltzmann}
\begin{figure}[h]
    \centering
    \includegraphics{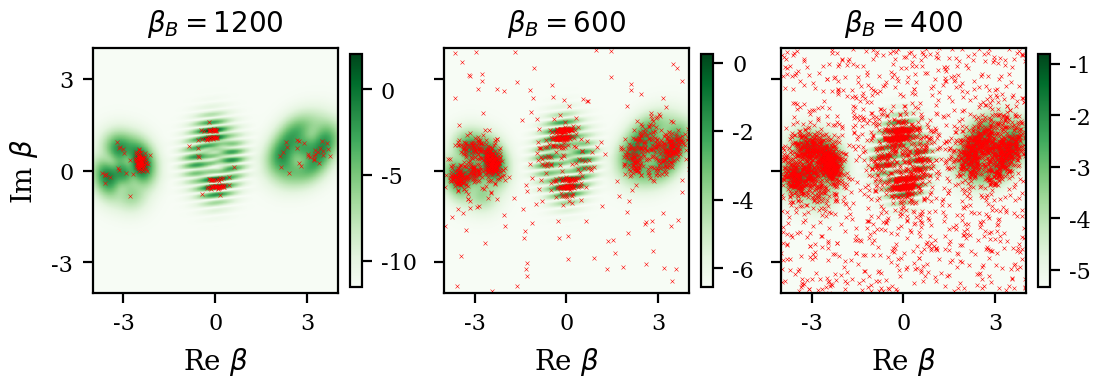}
    \caption{Probability density function of the Boltzmann distribution at different inverse temperatures $\beta_B$, together with the corresponding sampled points (red crosses). Larger $\beta_B$ values produce sharper distributions and concentrate sampling in more informative regions. Colour bars are logarithmic.}
    \label{fig:DevMapNew}
\end{figure}

As discussed in Sec.~\ref{subsec:Bootstrap}, the efficiency of sampling from the Boltzmann distribution (Eq.~\eqref{eq:Boltzmann}) strongly depends on the inverse temperature $\beta_B$, which controls the entropy of the distribution. The effect of $\beta_B$ on the resulting Boltzmann distribution is illustrated in Fig.~\ref{fig:DevMapNew}: increasing the inverse temperature sharpens the distribution, thereby reducing the probability of sampling points from uninformative regions of phase space. This improvement in sampling efficiency comes at the cost of increased sampling time, which in our implementation is determined by the rejection-sampling procedure~\cite{Bishop}.

\begin{figure}[h]
    \centering
    \includegraphics[]{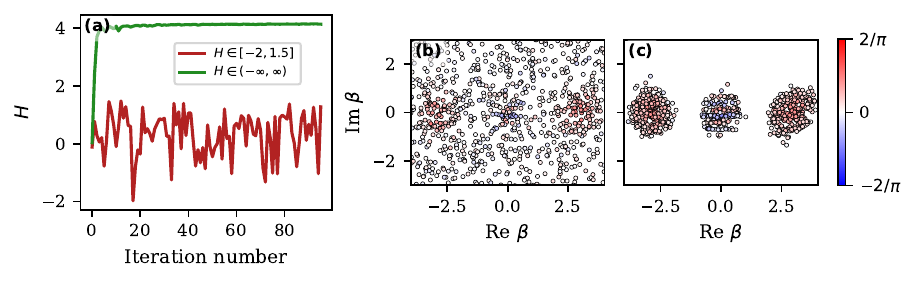}
     \caption{Entropy evolution during reconstruction for an adaptive strategy, in which the entropy $H$ is constrained to $[-2,1.5]$, and for a nonadaptive strategy without entropy adjustment, together with their impact on the sampling process. \textbf{(a)} Entropy as a function of the iteration number. \textbf{(b)} Sampled points for unconstrained sampling ($H \in (-\infty,\infty)$). \textbf{(c)} Sampled points for entropy-constrained sampling ($H \in [-2,1.5]$).} 
     \label{fig:entropy_over_time}
\end{figure}

To provide further intuition for the role of $\beta_B$ in the sampling process, we compare entropy-constrained sampling, restricted to the interval $[-2, 1.5]$ (corresponding to adaptive adjustment of $\beta_B$), to unconstrained sampling in Fig.~\ref{fig:entropy_over_time}. Figure~\ref{fig:entropy_over_time}\textbf{(a)} shows that, in the absence of entropy constraints, the entropy increases over time, corresponding to a progressive flattening of the sampling distribution. This behaviour is further illustrated by the distributions shown in Fig.~\ref{fig:entropy_over_time}\textbf{(b, c)}.

\subsection{Pipeline} 
\label{app:Pipeline}
Although all steps of the reconstruction algorithm are described in Sec.~\ref{sec:Reconstruction}, the practical implementation includes several modifications aimed at improving convergence and reducing reconstruction time.

\begin{figure}[h]
    \centering
    \includegraphics[width = \textwidth]{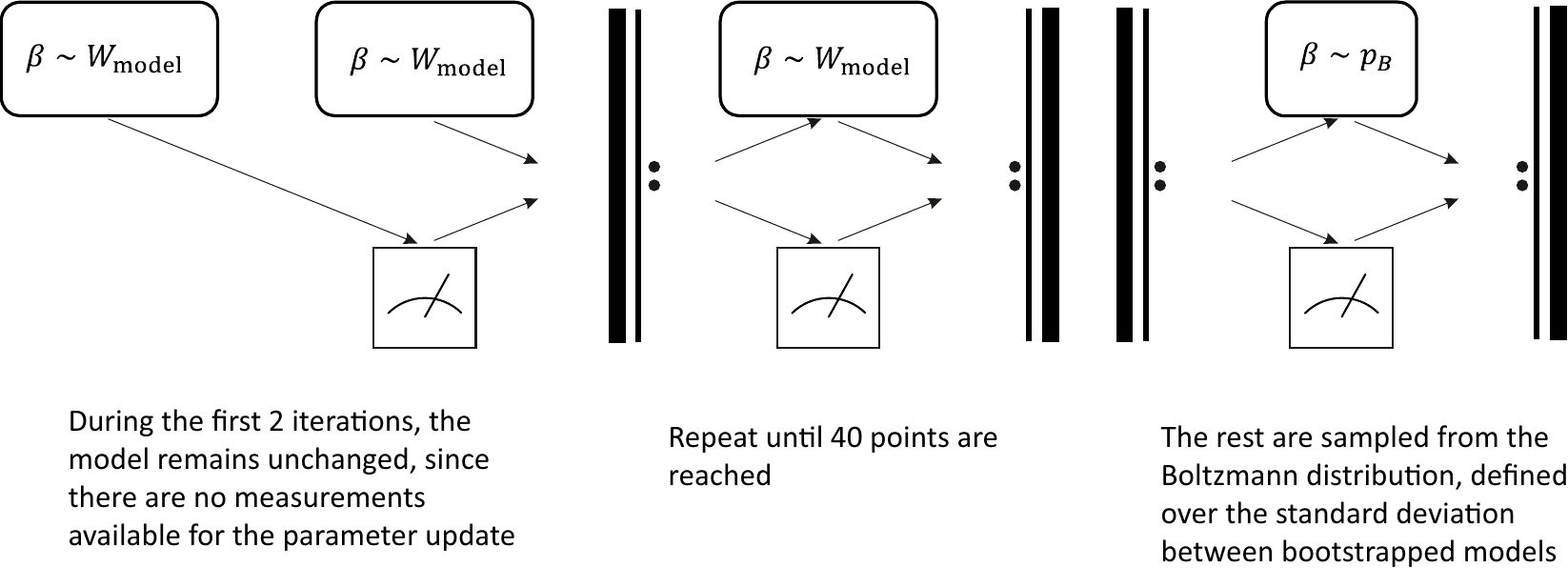}
    \caption{Reconstruction pipeline. Measurements and computation are performed in parallel threads. Initial iterations sample only from the Gaussian components of the initial model, while subsequent iterations may employ adaptive sampling strategies. Reconstruction terminates after a fixed number of iterations or once the variance across bootstrapped models falls below a threshold.}
    \label{fig:Pipeline}
\end{figure}

To increase reconstruction speed, hardware data acquisition and CPU-based calculations (i.e. MAP estimations for 10 bootstrapped models) for selecting subsequent measurement points are performed in parallel using two independent threads (see Fig.~\ref{fig:Pipeline}). In this configuration, the duration of a single iteration is determined by the longer of the two processes, measurement or computation. In the case when both the measurement and calculations require comparable time, this parallelisation yields an effective $\times 2$ speed-up. In this implementation, however, the selection of new sampling points is always delayed by one iteration as it relies on measurement data acquired in the previous step.

During the first two reconstruction iterations, sampling is restricted to the Gaussian components defined by the initial model guess (i.e. ideal cat state), since no measurement data are yet available to update the model parameters. From the third iteration onward, any sampling strategy may be employed. However, at the early stages of reconstruction, predictions from bootstrapped models may differ substantially or remain nearly identical, depending on the specific bootstrapped datasets and measurement outcomes. As a result, the optimal initial temperature for sampling (see Fig.~\ref{fig:DevMapNew}) varies significantly between the runs. To enhance stability, the first 40 sampling points (the number of iterations depends on the batch size) are always drawn from three Gaussians that define the model in Eq.~\eqref{eq:model}.

The reconstruction is terminated when either a predefined maximum number of iterations is reached or when the integrated variance across the ensemble of bootstrapped models falls below a specified threshold. The choice of these termination criteria is task- and parameter-dependent and is discussed in more detail in the context of the $\alpha = 3$ cat state reconstruction presented in the main text in Secs.~\ref{subsec:algorithm} and \ref{subsec:termination}.

\subsection{Optimal parameters} 
\label{app:OptimalParameters}
Based on the trade-off between reconstruction time and the variance of the reconstructed state fidelity, analysed in detail in Appendix~\ref{app:ReconstructorConvergenceCurves}, we select the set of reconstruction algorithm parameters used for the closed-loop optimisation runs in Sec.~\ref{sec:ClosedLoop}:
\begin{table}[h]
    \centering
    \begin{tabular}{c|c}
        batch size & 10 \\
         number of models & 10 \\
         random initialisation & False \\
         initial $\beta_{B}$ & 60 \\
         $\gamma$ & 0.25 \\
         amplitude imbalance & 0.05 \\
         $H$ & $[-2, 1.5]$ \\
         min \# of iterations & 25 \\
         max \# of iterations & 100
    \end{tabular}
    \caption{Reconstruction parameters used for closed-loop cat state optimisation. More details on the model initialisation (random vs using the model parameters from the previous iteration) are shown in Fig.~\ref{fig:true_vs_false}.}
    \label{tab:ReconstructionClosedLoop}
\end{table}
\section{Sampling strategies} \label{app:Sampling}
All the plots in Sec. \ref{sec:SamlingComparison} and this Section use the total number of measurements $n_\textrm{tot} = 2 \times n \times n_\textrm{avg}$, where $n$ is the number of measured points in phase space, $n_\textrm{avg}$ is the number of averages per measurement point, and a factor of 2 is coming from the background measurement for the Wigner function to eliminate cross-Kerr. Wigner function values for $p^\textrm{abs}$ and $p^\textrm{sq}$ sampling strategies are measured with $n_\textrm{avg} = 100$ averages per measurement point. Note that in Sec. \ref{sec:Reconstruction} all the plots use the number of measured points in phase space $n$, resulting in different axis scaling for plots in these two sections.

\subsection{Grid sampling} \label{app:GridSampling}
\begin{figure}[h!]
    \centering
    \includegraphics[]{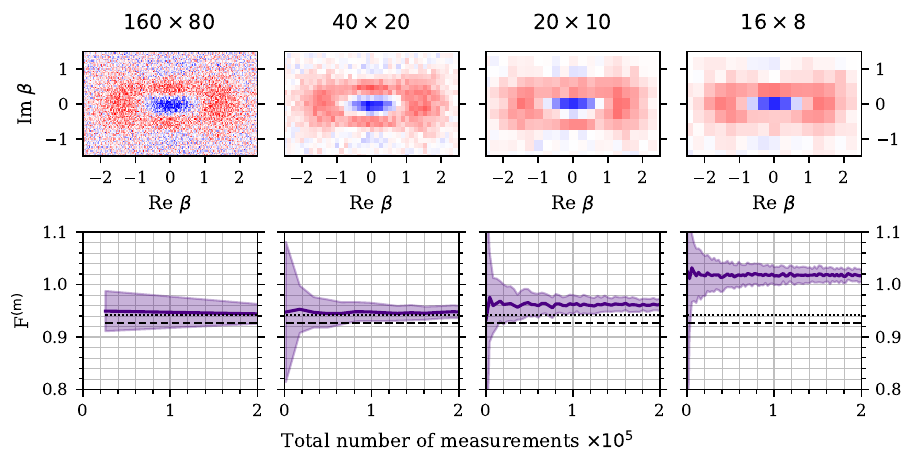}
    \caption{Influence of grid size on the fidelity estimate. Columns correspond to grids of size $n_I \times n_Q$, indicated at the top. The maximum number of measurements is fixed at $n_\mathrm{max} = 2 \cdot 10^5$. \textbf{First row:} Wigner maps for each grid, with the number of averages per point given by $\lfloor n_\mathrm{max}/(n_I \cdot n_Q) \rfloor$. \textbf{Second row:} Estimated fidelity between the measured and ideal states $F^\textrm{(m)}$ as a function of the total number of measurements $n_\mathrm{tot}$, obtained by fixing the grid and varying the number of averages per point up to $n_\mathrm{max}$. Dotted line: fidelity between the measured high-resolution Wigner map and the ideal cat state $F^\mathrm{(m)} = 0.94$. Dashed line: fidelity between the fit to the measured high-resolution Wigner map and the ideal cat state $F^\mathrm{(f)} = 0.93$. }
    \label{fig:grid_undersampling_ideal}
    \includegraphics[]{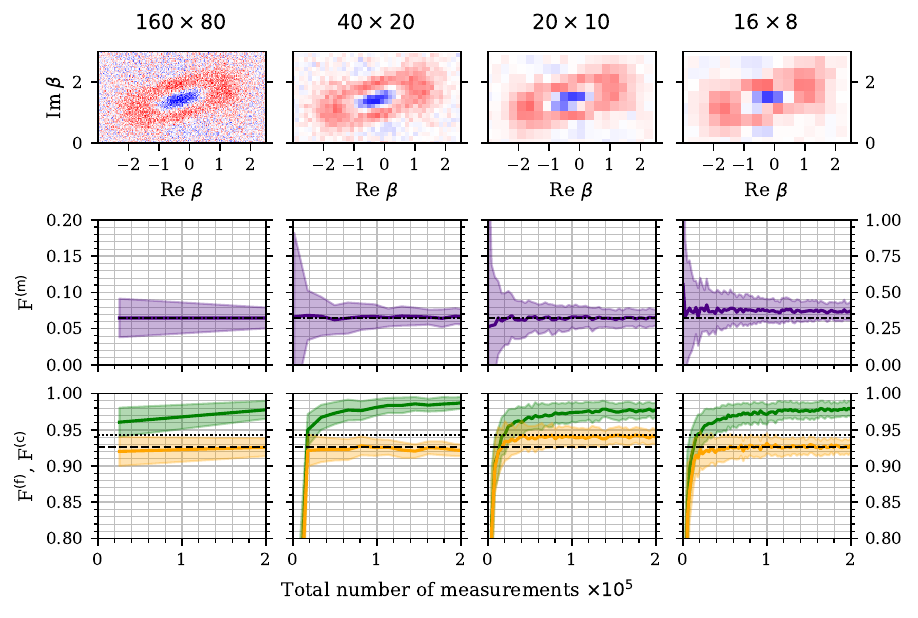}
    \caption{Influence of grid size on the fidelity estimate. Columns correspond to grids of size $n_I \times n_Q$, indicated at the top. The maximum number of measurements is fixed at $n_\mathrm{max} = 2 \cdot 10^5$. \textbf{First row:} Wigner maps for each grid, with the number of averages per point given by $\lfloor n_\mathrm{max}/(n_I \cdot n_Q) \rfloor$. \textbf{Second row:} Estimated fidelity between the measured and ideal state ($F^\textrm{(m)}$, green) as a function of the total number of measurements $n_\mathrm{tot}$, obtained by fixing the grid and varying the number of averages per point up to $n_\mathrm{max}$. Dash-dot line: fidelity between the measured high-resolution Wigner map and the ideal cat state $\lvert C_{\sqrt{2}}^{\pi}\rangle$. \textbf{Third row:} Estimated fidelity between the fit to measured Wigner values and the fit to a high-resolution Wigner map of the cat state ($F^\textrm{(f)}$, green), and fidelity between a corrected and ideal cat state ($F^\textrm{(c)}$, orange) as a function of the total number of measurements $n_\mathrm{tot}$, obtained by fixing the grid and varying the number of averages per point up to $n_\mathrm{max}$. Dotted and dashed lines as described in Fig.~\ref{fig:grid_undersampling_ideal}. }
    \label{fig:grid_undersampling_displ}
\end{figure}

To investigate the influence of undersampling the grid on the fidelity estimate, we measure the high-resolution grid with 800 averages and then downsample it to obtain rougher grids. We plot the estimated fidelity as a function of the total number of measurements $n_\textrm{tot}$ by increasing the number of averages per grid pixel, while keeping the grid fixed. We estimate the standard deviation by bootstrapping the measured data for each point. The Wigner maps for downsampled grids and fidelity plots corresponding to those grids are presented in ~\crefrange{fig:grid_undersampling_ideal}{fig:grid_undersampling_displ}.
\newpage
\FloatBarrier

\subsection{\texorpdfstring{$p^\textrm{abs}$}{|W|-sampling}} \label{app:AbsSampling}
The measurement setup used in this work supports high-power single-shot non-QND readout~\cite{Reed_2010}. However, owing to cross-Kerr between the cavity and the readout resonator, the relative number of shots associated with the qubit states $\lvert g\rangle$ and $\lvert e\rangle$ acquires a phase-space-dependent bias. This bias can be made independent of phase space position by applying a displacement of the opposite sign at the end of the Wigner sequence (Fig.~\ref{fig:overview}\textbf{(d)}), thereby keeping the cavity photon number constant during readout for all the Wigner measurements. While this procedure removes the phase space dependence, it does not eliminate the bias itself. In practice, the bias is corrected by performing a background Wigner measurement, implemented by applying positive and negative $\pi/2$ pulses at the end of the sequence, and averaging the results and subtracting them from each other (see Fig.~\ref{fig:overview}\textbf{(d)}). As a consequence, direct measurement of $p^\textrm{abs}(\beta)$ is not feasible in this setup.

\begin{figure}[h!]
    \centering
    \includegraphics[]{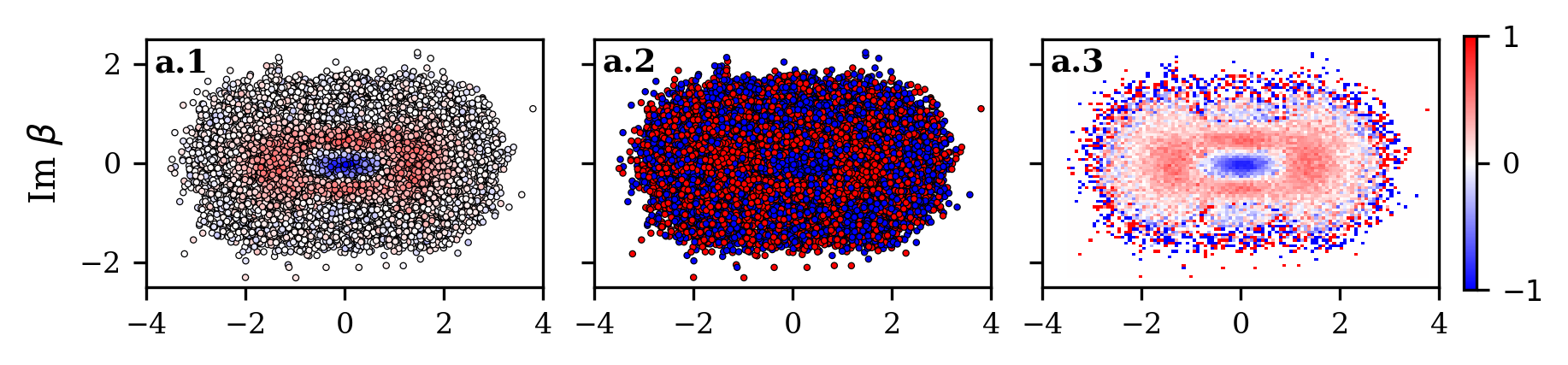} 
    \includegraphics[]{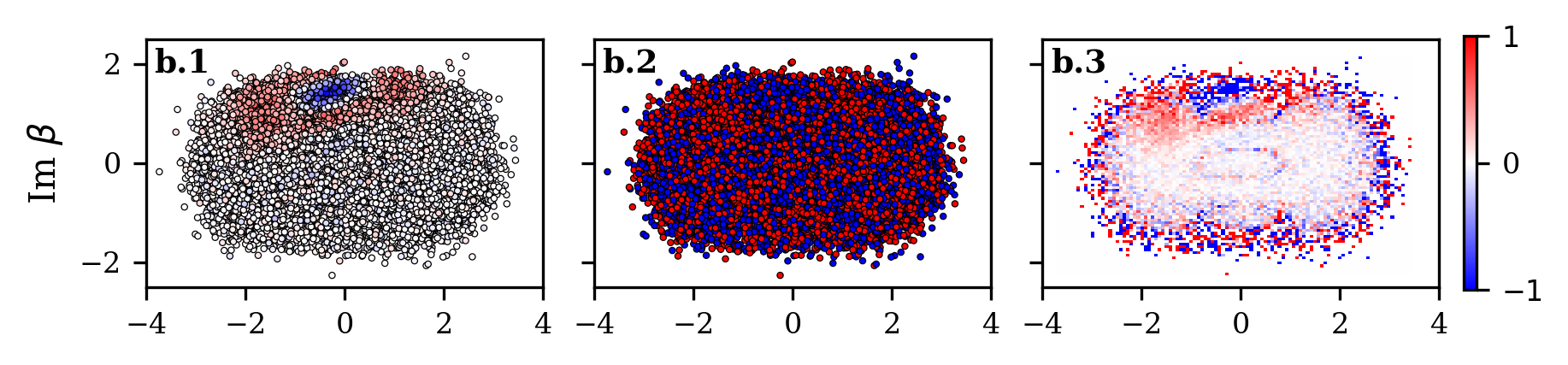}
    \caption{Artificial sampling of single-shot parity values based on the measured data. Plots \textbf{a.1 - a.3} correspond to the state $\lvert C_{\sqrt{2}}^\pi \rangle$, plots \textbf{b.1 - b.3} correspond to the state $R(15^\circ)D(1.5) \lvert C_{\sqrt{2}}^\pi \rangle$. \textbf{a.1 (b.1)} Parity, measured at points sampled from $p^\textrm{abs}(\beta)$. Each point is averaged 100 times. \textbf{a.2 (b.2)} Single-shot parity values, sampled using the Eq.~\eqref{eq:abs_parity_sim} and measurement data from \textbf{a.1 (b.1)}. \textbf{a.3 (b.3)} Parity map obtained from sampled parity data \textbf{a.2}. This map is not used for further evaluations and is plotted solely for clarity.}
    \label{fig:sampling_abs_steps}
\end{figure}

We bypass this issue as follows. We first sample $n=2 \cdot 10^5$ points from the distribution $p^\textrm{abs}(\beta)$. Then, we perform the Wigner measurement $W^+$ together with the Wigner background measurement $W^-$ for each point with 100 averages. This translates to the probability of finding the qubit in the excited state, $p_\mathrm{e} = 0.5 \cdot ((W^+ - W^-)\frac{\pi}{2} + 1)$. By using the obtained probability value $p_e$, we sample the parity value for each measured point in phase space
\begin{equation}
    P_\textrm{sim}(\beta) = 
    \left\{ \begin{aligned} 
  1, p = p_\mathrm{e},\\
  -1, p = 1 - p_\mathrm{e}.
\end{aligned} \right.
\label{eq:abs_parity_sim}
\end{equation}
This value is used for further calculations. The described data processing is illustrated in Fig. \ref{fig:sampling_abs_steps}. We follow \cite{LyonNN} for representing the resulting Wigner tomography based on $P_\textrm{sim}(\beta)$ at the points sampled from $\beta \sim p^\textrm{abs}(\beta)$ in Fig. \ref{fig:sampling_abs_steps} \textbf{a.3, b.3}. We construct a $101\times81$-bin histogram, and average $P_\textrm{sim}(\beta)$ values inside each bin. The bins with fewer data points appear noisier.

The suggested data processing requires measuring $2 \cdot 10^5$ points with high averaging and takes around 7 hours, making it sensitive to system drifts. For comparison, to collect $n_\textrm{tot} = 2 \cdot 10^5$ for $\beta \sim p^{sq}(\beta)$, we need to measure only $10^3$ points in phase space, which takes around 6 minutes. 
\newpage
\section{\texorpdfstring{Reconstruction convergence for $\alpha = 3$ cat state}{Reconstruction convergence for alpha = 3 cat state}}
\label{app:ReconstructorConvergenceCurves}

The reconstruction performance is benchmarked using an $\alpha = 3$ cat state. To enhance Kerr-induced distortions, the disentanglement pulse is intentionally chosen to be significantly longer than required and is implemented as a $4\sigma$ Gaussian with $\sigma = 40$ ns. Unselective qubit pulses are realised using $4\sigma$ Gaussian waveforms with $\sigma = 9$ ns.

\begin{figure}[h!]
    \centering
    \includegraphics{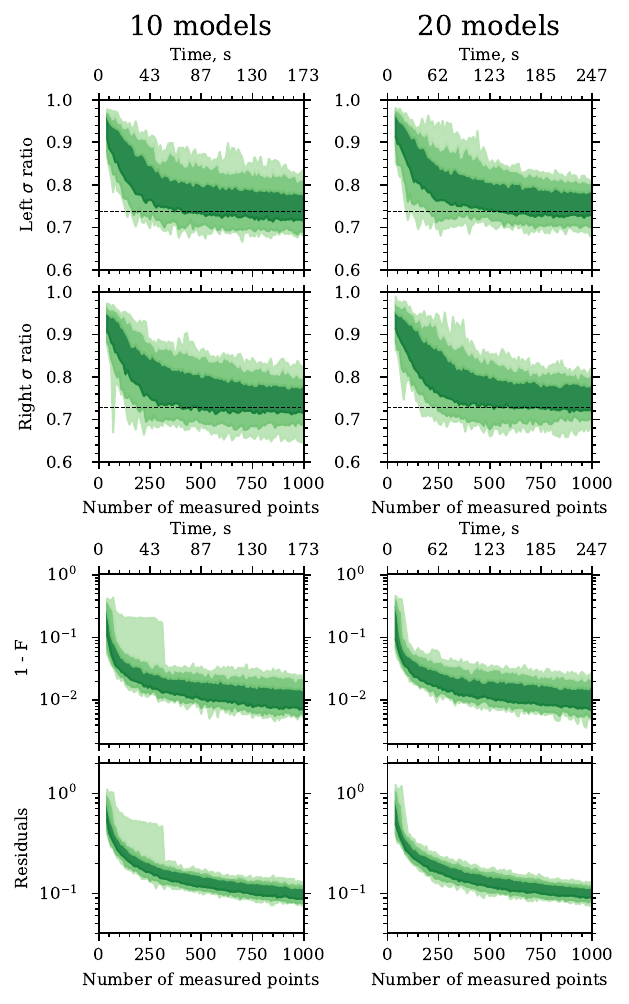}
    \caption{Convergence curves for reconstructions using 10 and 20 bootstrapped models. All runs employ rejection sampling with entropy constraint $H \in [-2,1.5]$, initial inverse temperature $\beta_B = 60$, 100 averages per point, and 10 measurements per batch.}
    \label{fig:10_vs_20_models}
\end{figure}

\begin{figure}[h!]
    \centering
    \includegraphics{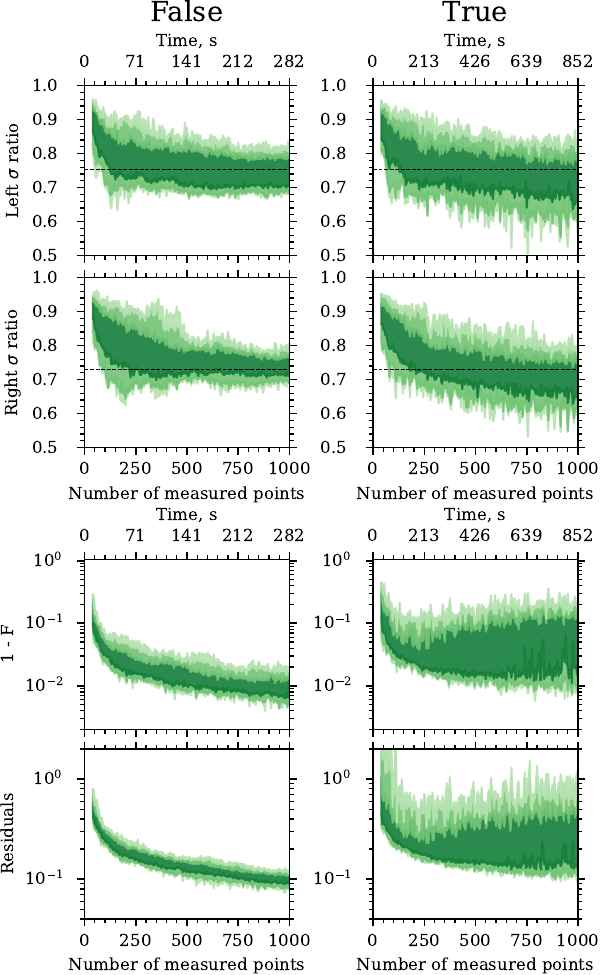}
    \caption{Convergence curves for different bootstrapped model initialisation strategies. The first column corresponds to initialisation using the model parameters from the previous iteration (random initialisation = False), while the second column uses parameters sampled from the prior distribution (random initialisation = True). All runs employ rejection sampling with entropy constraint $H \in [-2,0.1]$, 100 averages per point, and 5 measurements per batch. Each plot shows 32 independent runs for the corresponding configuration. }
    \label{fig:true_vs_false}
\end{figure}

The benchmarking explores the influence of the number of averages, batch size, and the number of bootstrapped models, as illustrated in \crefrange{fig:10_vs_100_vs_1000_avgs}{fig:10_vs_20_models}. In addition, we examine the impact of the initial parameter guess in MAP estimation on different reconstruction strategies, as shown in Fig.~\ref{fig:true_vs_false}.

\begin{figure}[h]
    \centering
    \includegraphics{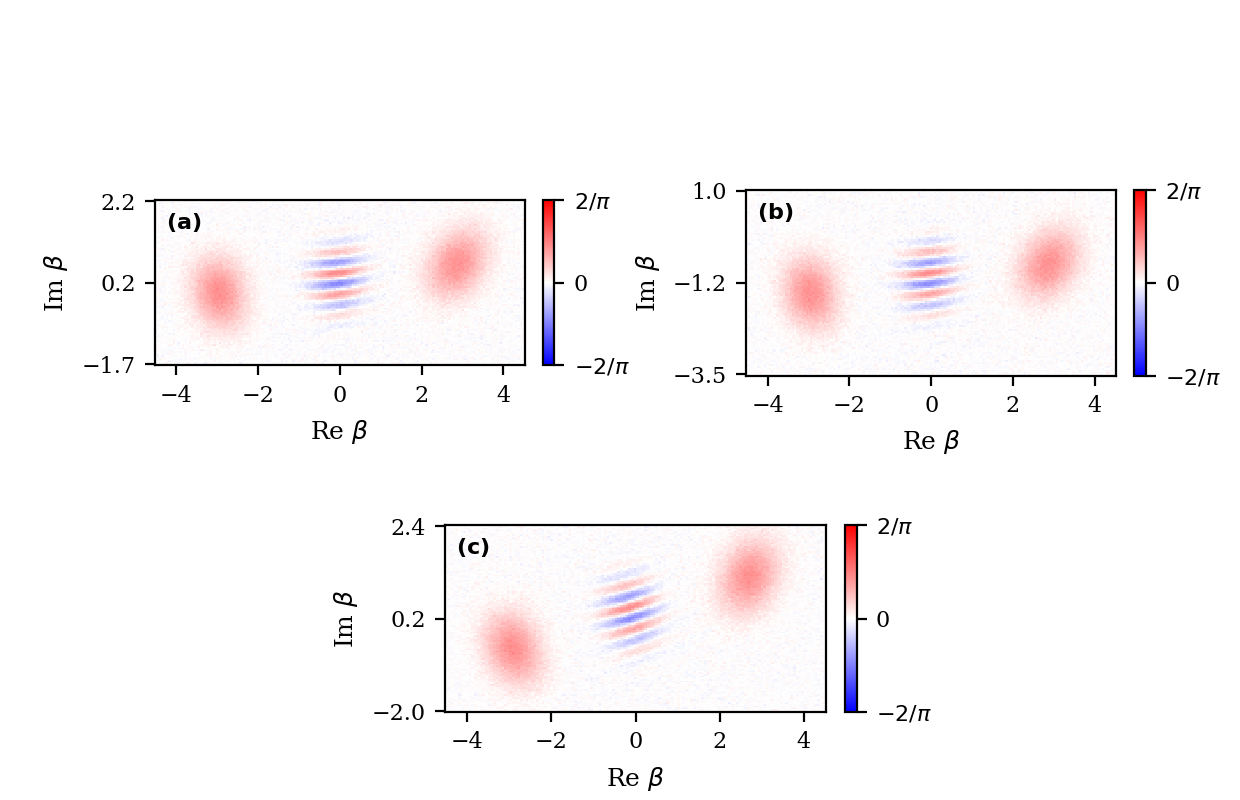}
    \caption{Cat states used for benchmarking the reconstruction method. All Wigner maps are measured with 1000 averages per point. \textbf{(a)} Cat state generated using the qcMAP sequence with a disentanglement pulse of duration $\sigma = 40$ ns. \textbf{(b)} Same state as in \textbf{(a)}, with the phase-space axes shifted by $(0,-1)$ in the measurement sequence. \textbf{(c)} Same state as in \textbf{(a)}, with the phase-space axes rotated by $10^\circ$ in the measurement sequence.}
    \label{fig:cat_vs_displaced_vs_rotated}
\end{figure}

\begin{figure}[h!]
    \centering
    \includegraphics{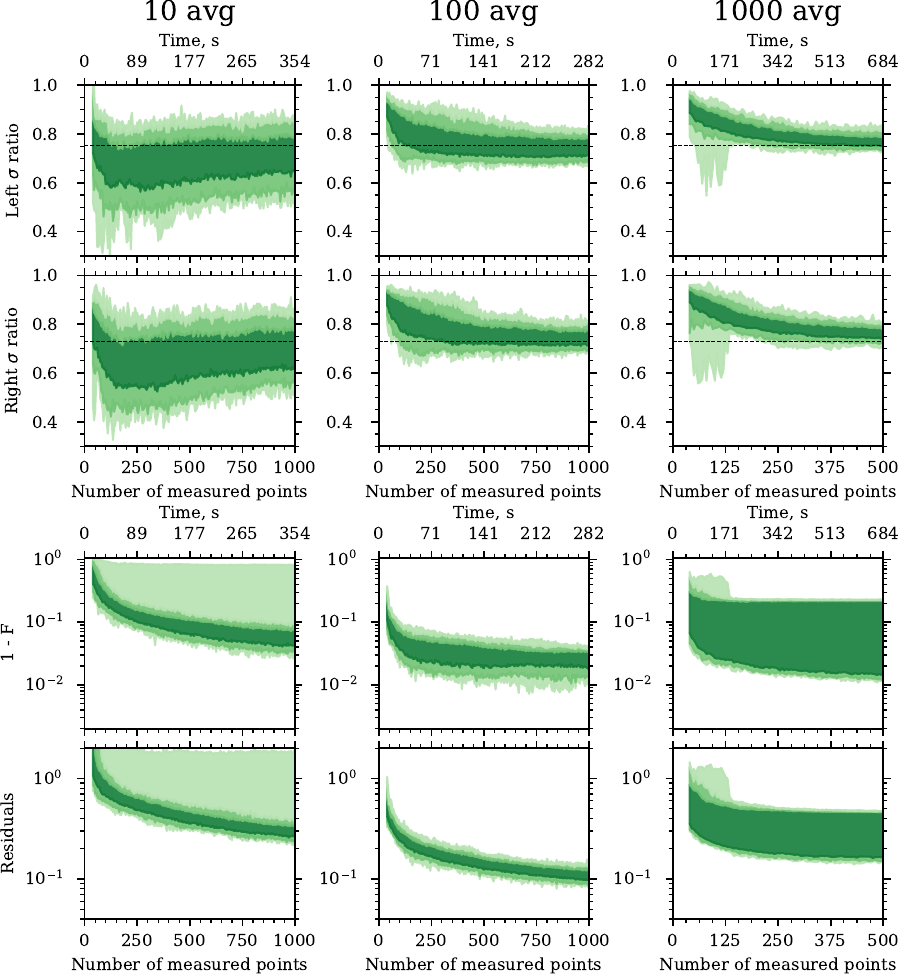}
    \caption{Convergence curves for different numbers of averages per measurement point. The left, centre, and right columns correspond to 10, 100, and 1000 averages, respectively. All runs employ rejection sampling with entropy constraint $H \in [-2,0.1]$ and initial inverse temperature $\beta_B = 60$, and reconstruct the uncorrected cat state (see Fig.~\ref{fig:cat_vs_displaced_vs_rotated}\textbf{(a)}). The batch size is 5.}
    \label{fig:10_vs_100_vs_1000_avgs}
\end{figure}

\begin{figure}[h!]
    \centering
    \includegraphics{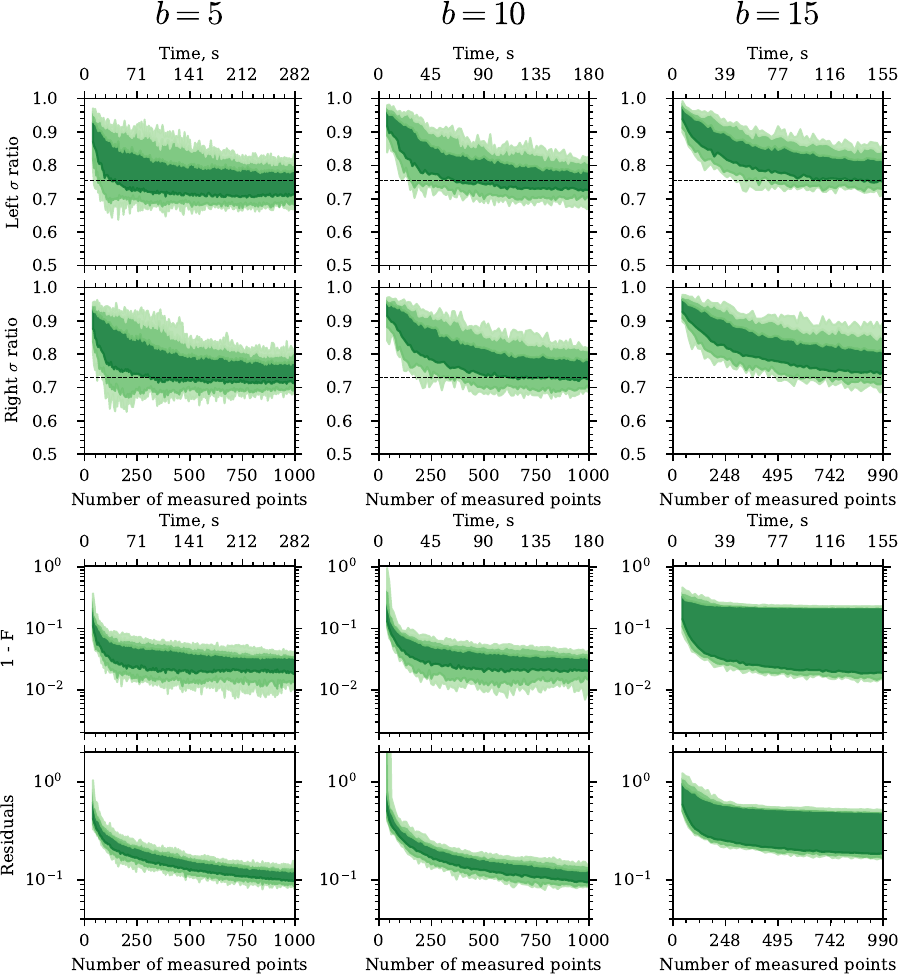}
    \caption{Convergence curves for different batch sizes, i.e. the number of measurement points acquired per reconstruction iteration. The left, centre, and right columns show 5, 10, and 15 samples per batch, respectively. All runs employ rejection sampling with entropy constraint $H \in [-2,0.1]$ and initial inverse temperature $\beta_B = 60$, and reconstruct the uncorrected cat state (see Fig.~\ref{fig:cat_vs_displaced_vs_rotated}\textbf{(a)}). Each measurement point is averaged 100 times.}
    \label{fig:5_vs_10_vs_15_bs}
\end{figure}

\begin{figure}[h!]
    \centering
    \includegraphics{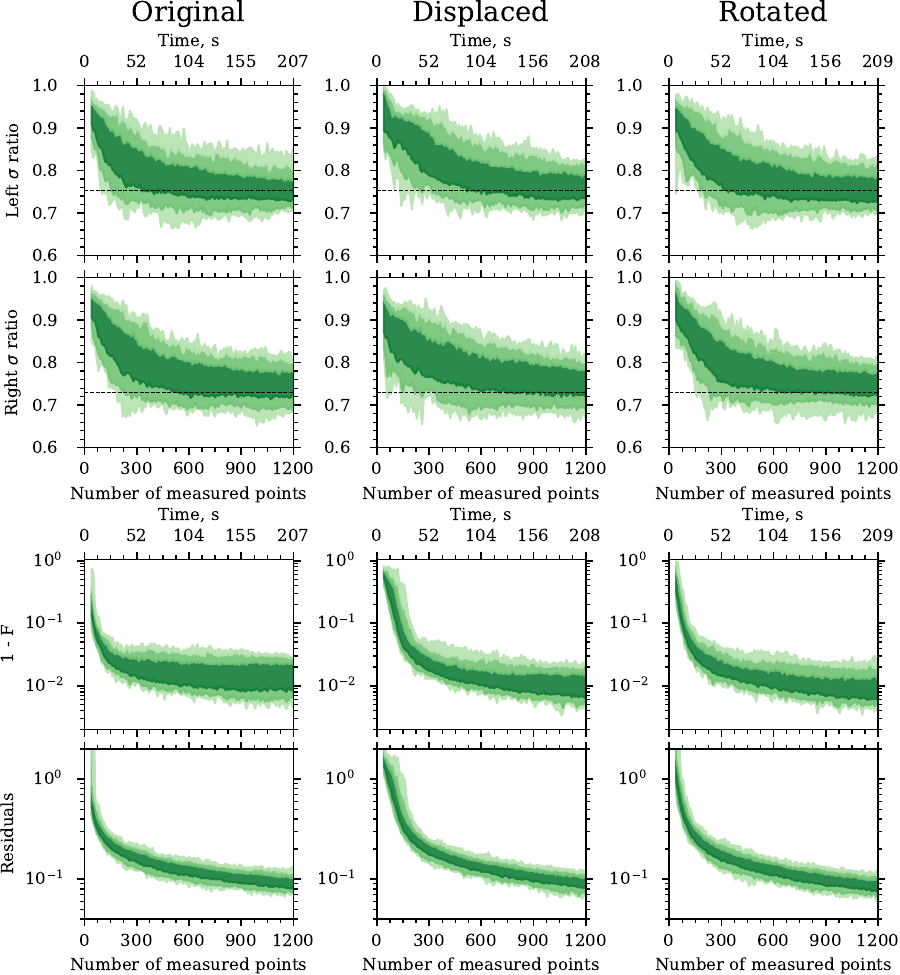}
    \caption{Convergence curves for the original, displaced, and rotated cat states. Reconstructions are performed with entropy constraint $H \in [-2,1.5]$, initial $\beta_B = 60$, 100 averages per point, and 10 measurements per batch. See Fig.~\ref{fig:cat_vs_displaced_vs_rotated} for the corresponding Wigner maps.}
    \label{fig:original_vs_displaced_vs_rotated}
\end{figure}

To further assess the convergence stability of the reconstruction algorithm, we performed reconstructions with a fixed set of hyperparameters for the original cat state as well as for artificially displaced and rotated states (see Fig.~\ref{fig:cat_vs_displaced_vs_rotated} for the corresponding Wigner maps). The results are shown in Fig.~\ref{fig:original_vs_displaced_vs_rotated}. Across all tested states, the reconstruction algorithm demonstrated robust, reproducible convergence, despite a significant mismatch between the prior parameter distributions and the initial parameter guess used for reconstruction and the target state parameters. This behaviour indicates that the reconstruction remains reliable even when the state undergoes large displacements or rotations in phase space.

\begin{figure}[h!]
    \centering
    \includegraphics{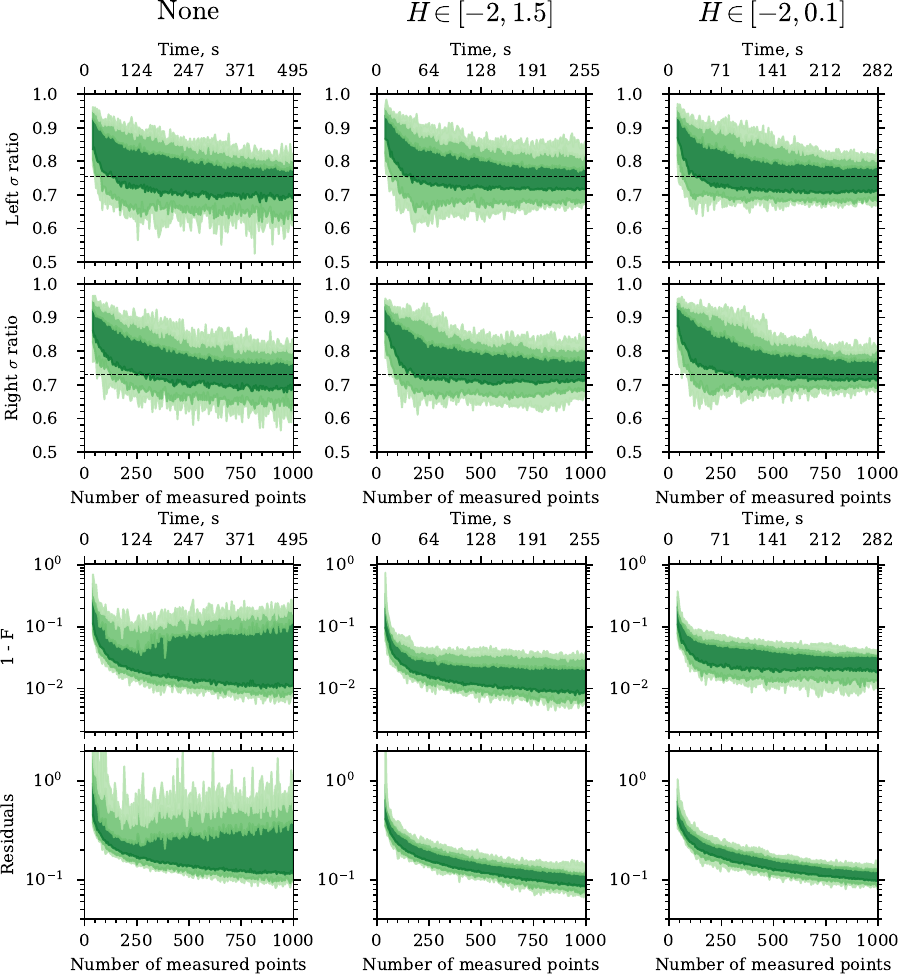}
    \caption{Convergence curves for different sampling strategies. The left column corresponds to sampling from the model distribution, the centre and right columns to rejection sampling with different allowed entropy ranges, $H \in [-2, 1.5]$ and $H \in [-2, 0.1]$. Each iteration samples and measures 5 points in phase space with 100 averages. For rejection sampling, the initial $\beta_B = 60$. All the runs reconstruct the uncorrected cat state (see Fig. \ref{fig:cat_vs_displaced_vs_rotated}\textbf{(a)}).}
    \label{fig:rejction_vs_no_rejection}
\end{figure}

\FloatBarrier
\section{Reconstruction convergence for smaller cat states} \label{app:MoreReconstruction}
\begin{figure}[h!]
    \centering
    \includegraphics{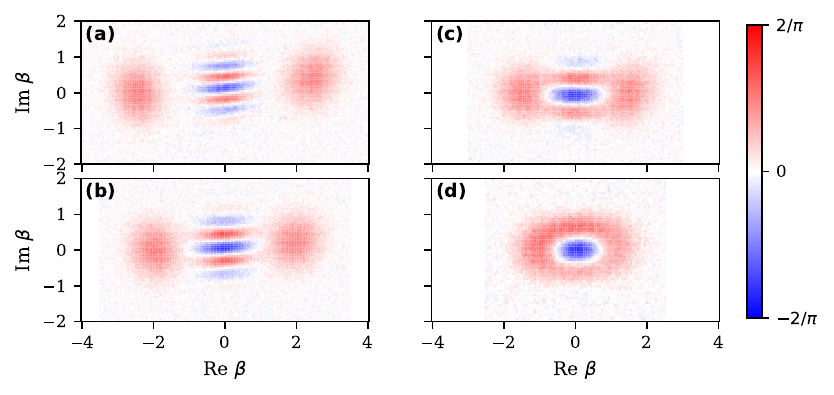}
    \caption{Wigner maps for cat states of different sizes: \textbf{(a)} $\alpha = 2.5$, \textbf{(b)} $\alpha = 2$, \textbf{(c)} $\alpha = 1.5$, \textbf{(d)} $\alpha = 1$. }
    \label{fig:smaller_cats_wigners}
\end{figure}

\begin{figure}[h!]
    \centering
    \includegraphics{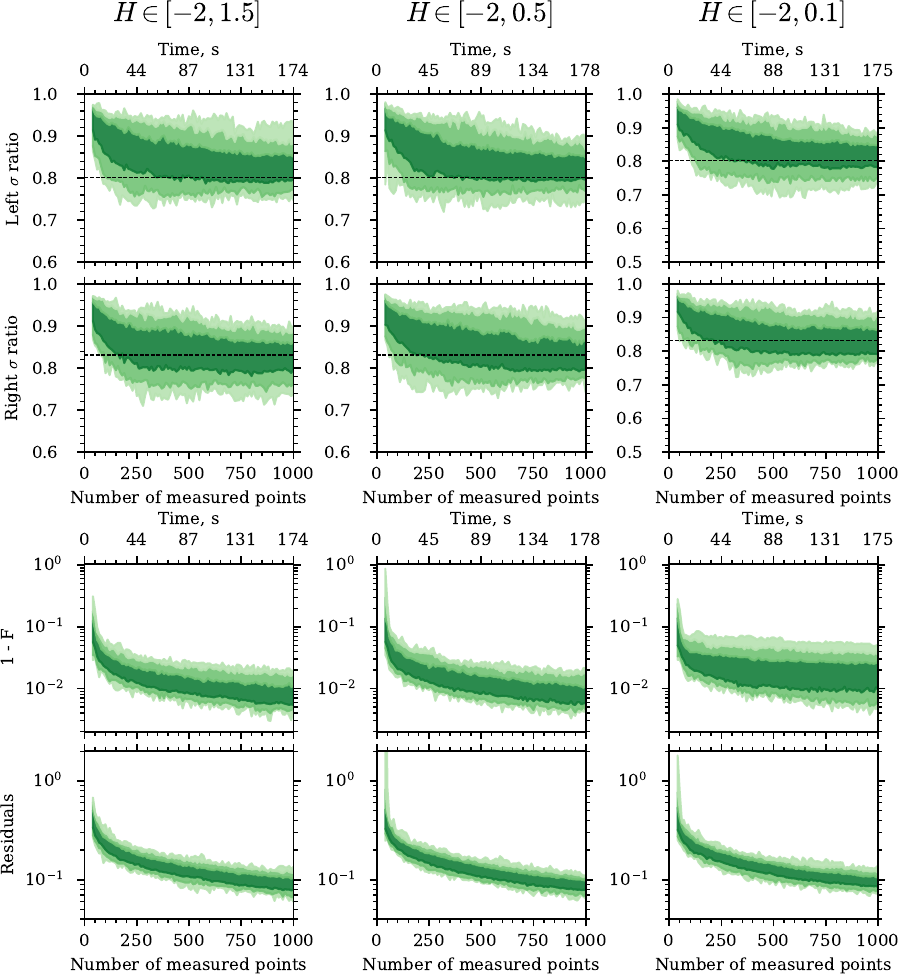}
    \caption{Convergence curves for different entropy ranges for $\alpha = 2.5$ cat state (see Fig.~\ref{fig:smaller_cats_wigners}\textbf{(a)}). The left column corresponds to $H \in [-2,1.5]$, while the centre and right columns correspond to $H \in [-2,0.5]$ and $H \in [-2,0.1]$, respectively. All runs use an initial inverse temperature $\beta_B = 60$, 100 averages per measurement point, and 10 measurements per batch.}
    \label{fig:cat2p5}
\end{figure}

\begin{figure}[h!]
    \centering
    \includegraphics{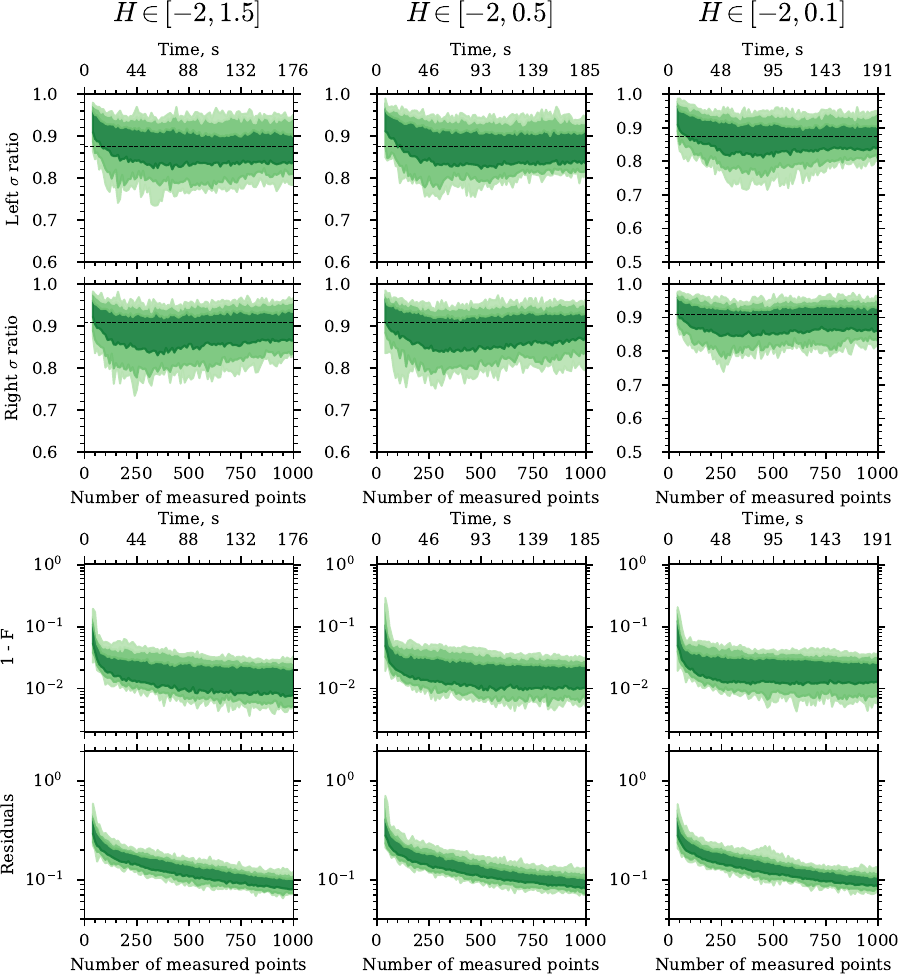}
    \caption{Convergence curves for different entropy ranges for $\alpha = 2$ cat state (see Fig.~\ref{fig:smaller_cats_wigners}\textbf{(b)}). The left column corresponds to $H \in [-2,1.5]$, while the centre and right columns correspond to $H \in [-2,0.5]$ and $H \in [-2,0.1]$, respectively. All runs use an initial inverse temperature $\beta_B = 60$, 100 averages per measurement point, and 10 measurements per batch.}
    \label{fig:cat2}
\end{figure}

\begin{figure}[h!]
    \centering
    \includegraphics{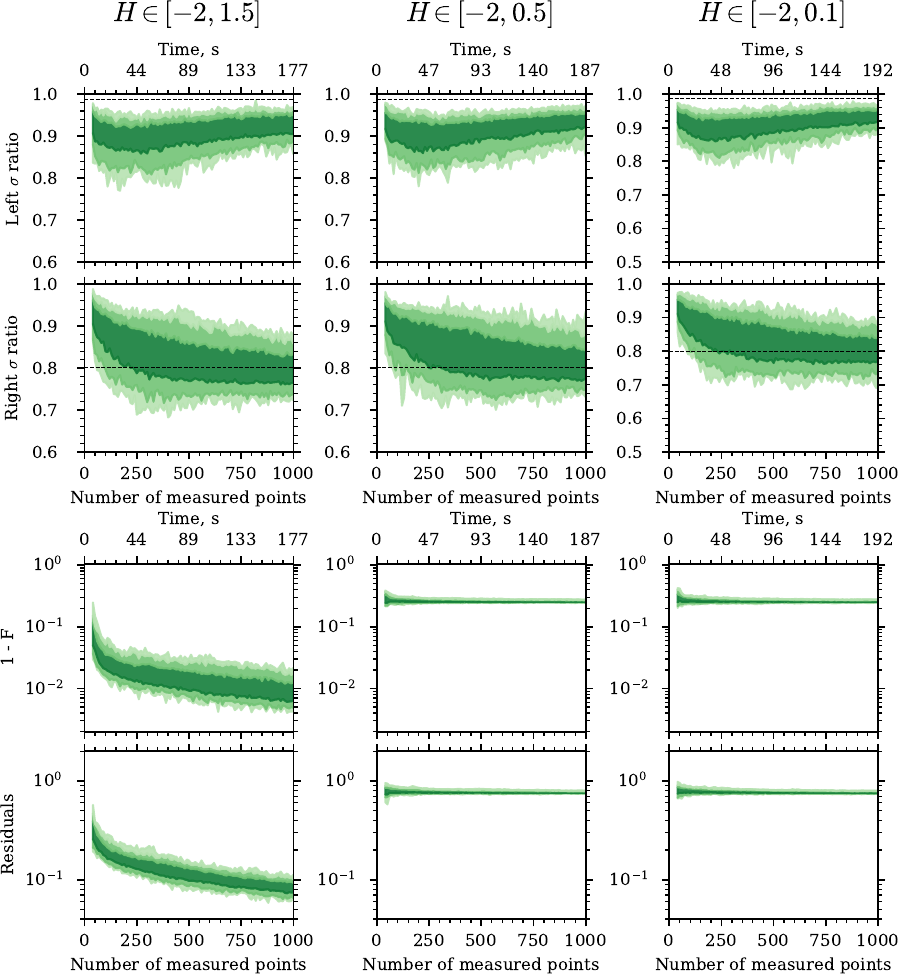}
    \caption{Convergence curves for different entropy ranges for $\alpha = 1.5$ cat state (see Fig.~\ref{fig:smaller_cats_wigners}\textbf{(c)}). The left column corresponds to $H \in [-2, 1.5]$, while the centre and right columns correspond to $H \in [-2, 0.5]$ and $H \in [-2, 0.1]$. All runs use an initial inverse temperature $\beta_B = 60$, 100 averages per measurement point, and 10 measurements per batch.}
   \label{fig:cat1p5}
\end{figure}

\begin{figure}[h!]
    \centering
    \includegraphics{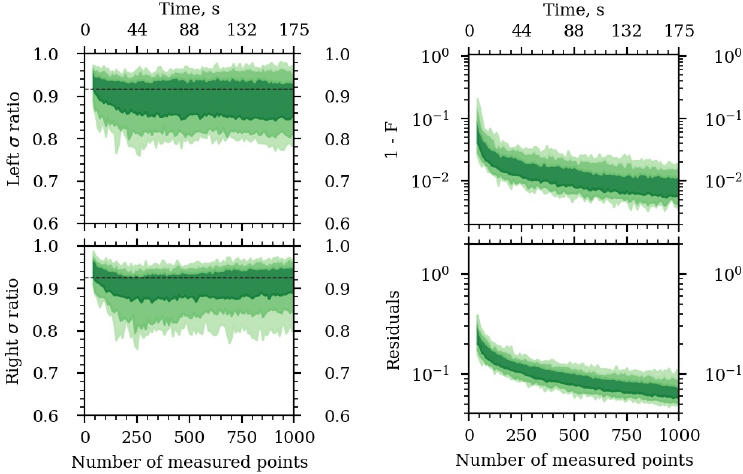}
    \caption{Convergence curves for $\alpha = 1$ cat state (see Fig.~\ref{fig:smaller_cats_wigners}\textbf{(d)}) reconstruction with rejection sampling, $H \in [-2, 1.5]$. For $\alpha = 1$ cat state, only one set of hyperparameters was characterised.}
   \label{fig:cat1}
    \includegraphics{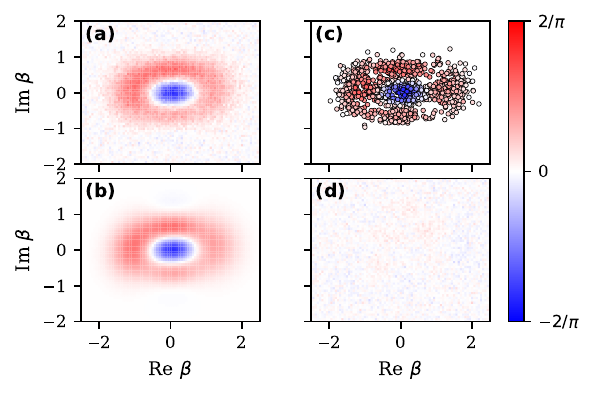}
    \caption{Example of a successful $\alpha = 1$ cat reconstruction for $H \in [-2, 1.5]$. \textbf{(a)} Reference high-resolution Wigner map. \textbf{(b)} Reconstructed Wigner map. \textbf{(c)} 1000 measured points during the reconstruction. \textbf{(d)} Residuals between the measured and reconstructed data $R(I, Q) = W_\mathrm{meas}(I, Q) - W_\mathrm{model}(I, Q)$.} 
\end{figure}

This section contains additional reconstruction data (\cref{fig:cat2p5,fig:cat2,fig:cat1p5,fig:cat1}) for smaller cat states shown in Fig.~\ref{fig:smaller_cats_wigners}. Interestingly, for $\alpha = 1.5$ cat, sampling with tight entropy ranges $[-2, 0.1]$ and $[-2, 0.5]$ results in $0.2 - 0.3$ reconstruction infidelity: the model predicts a wrong cat rotation angle. We suspect that this behaviour is caused by the phase boundary constraint (see Sec.~\ref{app:ModelParameters}), which was lifted later. An alternative explanation would be that smaller states require more sampling randomness to achieve low infidelity.

All the hyperparameters not specified under the plots are the same as for the $\alpha = 3$ cat state.

\FloatBarrier
\section{Experimental setup} 
\label{app:ExpSetup}
\begin{figure}[h]
    \centering
    \includegraphics[]{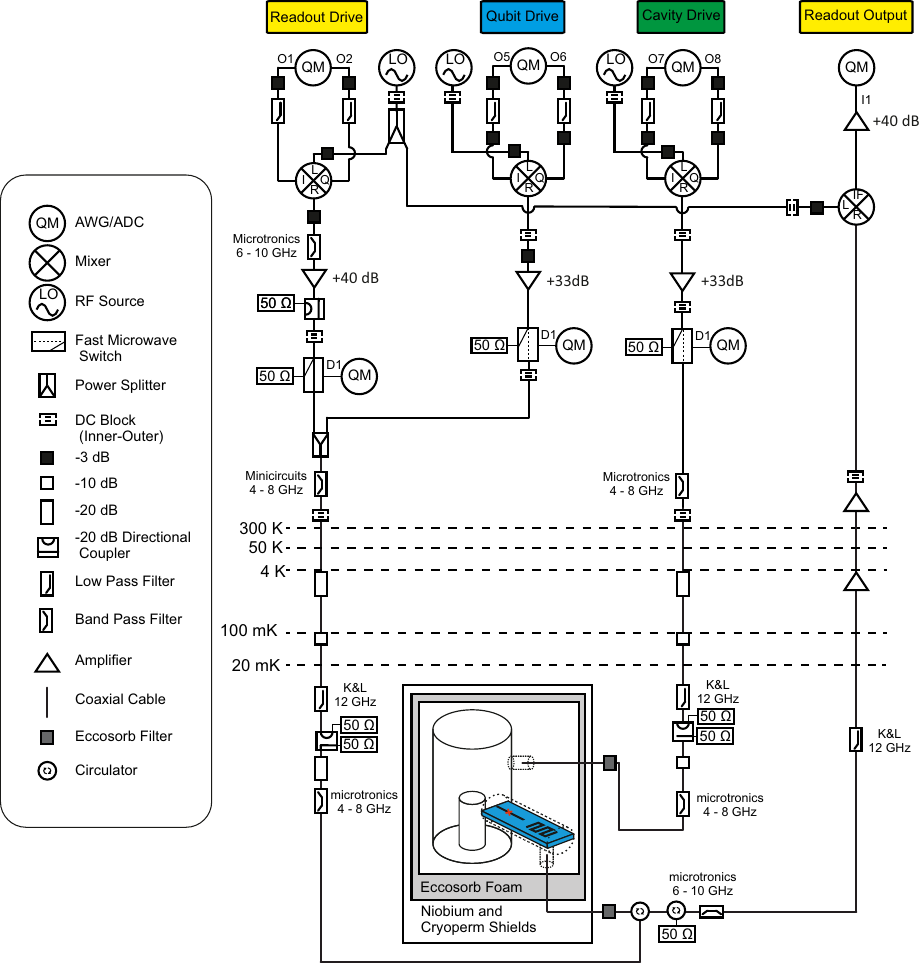}
    \caption{Experimental microwave setup. Inner-outer DC blocks are introduced in the setup to prevent ground loops, which affect the remaining flux-tunable experiments in the cryostat.}
    \label{fig:ExperimentalSetup}
\end{figure}
\begin{table}[h]
\centering
\begin{tabular*}{0.8\textwidth}{@{\extracolsep{\fill}}|p{5.2cm}|l|l|}
 \hline
 Parameter & Value & Measurement type \\
 \hline
 \hline
 Resonator frequency & $7.5284898(4)$ GHz 
 & spectroscopy\\
 \hline
 Resonator coupling $\kappa_c$ & $115.2 \pm 0.9$ kHz & spectroscopy \\
 \hline
 Resonator linewidth $\kappa_L$ & $148.2 \pm 1.6$ kHz & spectroscopy \\
 \hline
 Qubit ge frequency & $5.66936(9)$ GHz & spectroscopy \\
 \hline
 Qubit ef frequency & $5.479845(7)$ GHz& spectroscopy\\
 \hline
 Qubit anharmonicity & $189.5 \pm 0.1$ MHz& calculated (spectroscopy) \\
 \hline
 Qubit temperature, mK & $149.1 \pm 1.5$ mK & g-e and e-f Rabi\\
 \hline
 Qubit $T_1$ & $31.6 \pm 0.2 \ \mu s$& $T_1$ measurement\\
 \hline
 Qubit $T_2$ & $12.3 \pm 0.2 \ \mu s$& Ramsey\\
 \hline
 Qubit $T_2$ Echo & $14.9 \pm 0.4 \ \mu s$ & Echo measurement\\
 \hline
 Cavity frequency  & $4.5444444(6)$ GHz& spectroscopy\\
\hline
Cavity $T_1^c$ &  $59.0 \pm 0.7 \ \mu s$ & cavity $T_1$ measurement \\  
 \hline
 Cavity self-Kerr $K/2$ &  $4.6 \pm 0.1$ kHz & cavity revivals\\
\hline
Cavity thermal population $n_{th}$ & $0.071 \pm 0.004$ & displacement calibration \\
\hline
Cavity displacement scaling & $14.41 \pm 0.03$ & displacement calibration \\
\hline
Qubit-cavity dispersive shift $\chi$ & $1.514 \pm 0.001$ MHz & qubit revivals \\
\hline
\shortstack[c]{\rule{0pt}{2.6ex} Second order qubit-cavity \\ dispersive shift $\chi_{qcc}/2$}
&
\shortstack[c]{$11.6 \pm 1.0$ kHz\\$11.2 \pm 0.4$ kHz}
&
\shortstack[c]{qubit revivals\\cavity revivals}
\\
\hline
\end{tabular*}
\caption{Experimentally characterised Hamiltonian parameters of the system.}
\label{table:experimental_params_table}
\end{table}
\FloatBarrier
\section{Fisher Information and Cramér--Rao Bound}
\label{app:fisher-information}

The goal of this section is to quantify the fundamental precision limit on
fidelity estimation from Wigner function measurements of cat states for different sampling strategies. 
Given a finite total measurement budget $N_{\rm total}$, we derive lower bounds
on the statistical uncertainty $\sigma_F$ of any unbiased fidelity estimator
for each of the three sampling strategies: grid sampling and optimal strategies with $W^2$ and $|W|$ weighting.
All results use the constrained model introduced in
Sec.~\ref{subsec:WignerModel} and Appendix~\ref{app:ModelParameters}.
Throughout this section, $\boldsymbol{\eta} = (\eta_1,\ldots,\eta_{16})$ denotes the 16 free parameters
of Eq.~\eqref{eq:set_of_model_parameters}.

The derivation proceeds in four steps.
First, we write down the Bernoulli likelihood for a single displaced-parity measurement and use it to define the per-pixel Fisher information matrix (Sec.~\ref{sec:fim_likelihood}).
Second, we specialise it to the Wigner tomography measurement model and take the continuum limit, obtaining the Fisher density $\mathcal{M}$ (Sec.~\ref{sec:fim_measurement_model}).
Third, we contract $\mathcal{M}$ with the fidelity gradient via the Cram\'er--Rao bound to arrive at a lower bound on $\sigma_F$ and define the asymptotic prefactor $C_\infty$ (Sec.~\ref{sec:crb_fidelity}).
Finally, we evaluate $C_\infty$ semi-analytically for the three sampling strategies using the density $\mathcal{M}$ derived in step two, and compare the resulting fidelity uncertainty bounds (Sec.~\ref{sec:fim_analytics}). We then validate these analytic predictions numerically and visualise $\sigma_F$ as a function of $N_{\rm total}$ and $\alpha_0$ for all four strategies (Sec.~\ref{sec:fim_comparison}).

For the Fisher-information calculation, it is convenient to reparametrise the Gaussians
standard deviations $\sigma_{k}^{(j)}$ via logarithmic deviations: 
\begin{equation}
\label{eq:log_width_def}
\ell_{j,k} := \log\sigma_{k}^{(j)},
\qquad j\in\{L,R,C\},\quad k\in\{x,y\},
\end{equation}
so that $\frac{\partial}{\partial\ell_{j,k}} = \sigma_{k}^{(j)}\,\frac{\partial}{\partial\sigma_{k}^{(j)}}$.
This reparametrisation is volume-preserving ($d\ell_{j,k} = d\sigma_k^{(j)}/\sigma_k^{(j)}$,
so the Jacobian is unity) and renders the Gaussian gradient
formulas algebraically compact: the chain rule
$\partial/\partial\ell_{j,k} = \sigma_k^{(j)}\,\partial/\partial\sigma_k^{(j)}$
absorbs the prefactors that would otherwise appear in every derivative; the six $\ell_{j,k}$ replace the six
$\sigma_{k}^{(j)}$ in the parameter vector $\boldsymbol{\eta}$ while all other parameters remain as
in Eq.~\eqref{eq:set_of_model_parameters}.
Here $j \in \{L, C, R\}$ labels the left, central fringe, and right
Gaussian components respectively, and $k \in \{x, y\}$ labels the deviations along
the $I$- and $Q$-axes; each component therefore has two independent standard
deviations $\sigma^{(j)}_{x}$ and $\sigma^{(j)}_{y}$.
Furthermore, we write $\partial_a \equiv \partial/\partial\eta_a$ throughout this section.

\subsection{Likelihood function and Fisher information}
\label{sec:fim_likelihood}

Each measurement at phase-space pixel $\beta_i$ is a single-shot displaced
parity measurement with outcome $m \in \{+1,-1\}$.  The outcome probability
is
\begin{equation}
\label{eq:bernoulli_likelihood}
P(m \mid \beta_i;\boldsymbol{\eta}) = \frac{1 + m\,\tfrac{\pi}{2}\,W(\beta_i;\boldsymbol{\eta})}{2}.
\end{equation}
For brevity, in this section we use $W(\beta_i ;\,\boldsymbol{\eta})=W_\mathrm{model}(\beta_i ;\,\boldsymbol{\eta})$, which is defined in Eq.~\eqref{eq:model} in the main text. Generally, $\beta_i = I_i + iQ_i$,
and $I$, $Q$ denote the real and imaginary parts of the phase space coordinate
$\beta = I + iQ$ introduced in Sec.~\ref{subsec:WignerModel}.
After $N_i$ independent shots at pixel $i$, the log-likelihood is
\begin{equation}
\label{eq:log_likelihood}
\log\mathcal{L}(\boldsymbol{\eta})
= \sum_i \bigl[
  k_i \log P(+1\mid\beta_i;\boldsymbol{\eta})
  + (N_i - k_i)\log P(-1\mid\beta_i;\boldsymbol{\eta})
\bigr],
\end{equation}
where $k_i$ is the number of $+1$ outcomes.
The Fisher information matrix (FIM) is defined as minus the expected
Hessian of the log-likelihood,
\begin{equation}
\label{eq:fim_def}
\mathcal{I}_{ab}(\boldsymbol{\eta})
= -\mathbb{E}\!\left[
  \frac{\partial^2 \log\mathcal{L}}{\partial\eta_a\,\partial\eta_b}
\right],
\end{equation}
and the Cramér--Rao bound (CRB) states that the covariance matrix of any
unbiased estimator $\hat{\boldsymbol{\eta}}$ for a parameter vector $\boldsymbol{\eta}$ satisfies
$\mathrm{Cov}[\hat{\boldsymbol{\eta}}] \succeq \mathcal{I}(\boldsymbol{\eta})^{-1}$.
The scalar CRB on fidelity uncertainty is derived in Sec.~\ref{sec:crb_fidelity}.

\subsection{Wigner measurement model}
\label{sec:fim_measurement_model}

Wigner tomography is performed by displaced-parity measurements.
Evaluating Eq.~\eqref{eq:fim_def} with the Bernoulli likelihood of
Eq.~\eqref{eq:bernoulli_likelihood} gives
\begin{equation}
\label{eq:fim_wigner}
\mathcal{I}_{ab}(\boldsymbol{\eta})
= \sum_{i} \frac{N_i\,\pi^2\,\partial_a W_i\,\partial_b W_i}{4 - \pi^2 W_i^2},
\end{equation}
where $W_i \equiv W(\beta_i;\boldsymbol{\eta})$ and $N_i$ is the number of shots at
measurement point $i$.
For a sampling strategy with a fixed number of shots per point $N_{\rm shots}$,
the total measurement budget is $N_{\rm total} = N_{\rm shots}\cdot M$, where
$M$ is the number of phase space points, and
\begin{align}
\hat{\mathcal{M}}_{ab} &\equiv \frac{\mathcal{I}_{ab}(\boldsymbol{\eta})}{N_{\rm total}}
= \frac{1}{M} \sum_{i}
\frac{\pi^2\,\partial_a W_i\,\partial_b W_i}{4 - \pi^2 W_i^2}
\xrightarrow{M\to\infty} \mathcal{M}_{ab},
\label{eq:M_hat_ab}\\
\mathcal{M}_{ab} &= \mathbb{E}_{p(\beta)}\!\left[
\frac{\pi^{2}\,\partial_{a}W(\beta)\,\partial_{b}W(\beta)}
{4-\pi^{2}W(\beta)^{2}}\right],
\label{eq:M_ab}
\end{align}
where $p(\beta)$ denotes the probability density of the chosen sampling
distribution.
For the grid strategy, $M\to\infty$ at fixed $N_{\rm shots}$ is equivalent
to $\delta\to 0$; for the optimal strategies, it corresponds to drawing more
i.i.d.\ points from $p(\beta)$, and convergence follows from the law of
large numbers.
Here, $\delta$ is the pixel spacing between adjacent measurement points in
phase space, and the limit $\delta \to 0$ corresponds to the
Nyquist-resolved regime $\delta < \pi/(4\alpha_0)$, in which the discrete sum
converges to the integral $\mathcal{M}_{ab}$ independently of grid size.
For finite pixel spacing $\delta$, the estimator $\hat{\mathcal{M}}_{ab}$ is biased;
unbiasedness is only recovered once the Nyquist condition is satisfied.
Note that $\delta$ does not appear explicitly on the left-hand side of
Eq.~\eqref{eq:M_hat_ab} because, for the grid strategy, each pixel carries
$N_{\rm shots}$ shots so that $N_{\rm total} = N_{\rm shots}\cdot(\text{Area}/\delta^2)$;
the pixel spacing therefore enters through the total number of pixels rather than
through the per-pixel weights.

\subsection{Cram\'er--Rao bound for fidelity estimation}
\label{sec:crb_fidelity}
The fidelity between the parametrised state $\rho(\boldsymbol{\eta})$ and the target
cat state $\rho_{\rm cat}$ is estimated using the Wigner overlap defined in Eq.~\eqref{eq:fidelity}
\begin{equation}
\label{eq:fidelity_def}
F(\boldsymbol{\eta}) = \pi \int W_{\rm cat}(\beta)\,W(\beta;\boldsymbol{\eta})\,d^2\beta,
\end{equation}
where $W_{\rm cat}(\beta)=W_\mathrm{qcMAP}(\beta)$ defined in Eq.~\eqref{eq:qcmap_thermal_cat_wigner} with unit purity $\mathcal{P}=1$.
The corresponding fidelity estimator is obtained by evaluating this function at the parameter estimator, $\hat{F}=F(\hat{\boldsymbol{\eta}})$.
By the Cram\'er--Rao bound, the variance of any unbiased estimator $\hat{F}$
satisfies
\begin{equation}
\label{eq:crb_fidelity}
\sigma_F^2 \;\ge\; g_F^\top \mathcal{I}(\boldsymbol{\eta})^{-1} g_F,
\qquad g_F = \nabla_\eta F,
\end{equation}
where $\mathcal{I}$ is the FIM with components defined in Eq.~\eqref{eq:fim_wigner}
and the gradient components are $g_{F,a}=\partial F/\partial\eta_a$.
By introducing the estimator $\hat{\mathcal{M}}$ of the Fisher information density
$\mathcal{M}$ (see Eq.~\eqref{eq:M_ab}), Eq.~\eqref{eq:crb_fidelity} can be
rewritten as
\begin{equation}
\label{eq:sigma_from_M}
\sigma_F \sqrt{N_{\rm total}} \;\geq\;
\sqrt{g_F^\top \hat{\mathcal{M}}^{-1} g_F}.
\end{equation}

Equation~\eqref{eq:sigma_from_M} provides a practical procedure for evaluating
the Cram\'er--Rao lower bound on the fidelity uncertainty.
First, measurement points are generated according to the chosen sampling strategy
(grid, $|W|$, or $W^2$); the adaptive strategy is compared numerically to these
three in Sec.~\ref{sec:fim_comparison}.
Second, the Fisher information density $\hat{\mathcal{M}}$ is estimated from the
sampled phase space points using Eq.~\eqref{eq:M_hat_ab}.
Third, the fidelity gradient $g_F$ is evaluated (analytical expressions for the
ideal cat state are derived in Sec.~\ref{sec:fim_analytics}).
Finally, the lower bound on the fidelity uncertainty is obtained from
Eq.~\eqref{eq:sigma_from_M}.

As $N_{\rm total}\to\infty$, $\hat{\mathcal{M}}\to\mathcal{M}$, which motivates
the definition of the asymptotic prefactor
\begin{equation}
\label{eq:Cinf_def}
C_\infty := \sqrt{g_F^\top \mathcal{M}^{-1} g_F},
\end{equation}
so that $\sigma_F\sqrt{N_{\rm total}} \to C_\infty$ for resolution-independent
strategies, or $\sigma_F\sqrt{N_{\rm shots}}/\delta \to C_\infty$ for the grid sampling (see Table~\ref{tab:crb_prefactors}).
Physically, $C_\infty$ is the sampling strategy-dependent coefficient that sets
the fundamental lower bound on fidelity uncertainty in the high-statistics limit:
a smaller $C_\infty$ indicates that more Fisher information is extracted per
measurement, yielding a tighter bound on $\sigma_F$. This comparison is meaningful only between strategies that share the same
normalisation (see Table~\ref{tab:crb_prefactors}).
The grid normalisation $\sigma_F\sqrt{N_{\rm shots}}/\delta$ and the
optimal strategy normalisation $\sigma_F\sqrt{N_{\rm total}}$ are inequivalent, so $C_{\infty,\rm grid}$ and
$C_{\infty,W^2}$ or $C_{\infty,|W|}$  cannot be directly compared to rank the strategies.
The strategies with different normalisation can be compared by $\sigma_F$ at a fixed
$(\alpha_0, N_{\rm total})$. As shown later in Sec.~\ref{sec:fim_comparison}, at the reference point
$(\alpha_0,N_{\rm total})=(3,10^5)$ with $N_{\rm shots}=40$, the grid strategy gives
$\sigma_F^{\rm grid}\approx0.021$ while the optimal strategies give
$\sigma_F^{W^2}\approx0.006$ and $\sigma_F^{|W|}\approx0.007$ -- a factor
of $\sim3$ improvement (see Fig.~\ref{fig:uncertainity_sampling_strategies}).

The value of $C_\infty$ depends on the sampling strategy and is derived
analytically for all three cases in the next subsection.

\subsection{\texorpdfstring{Semi-Analytic derivation of $C_\infty$}{Semi-Analytic derivation of C}}
\label{sec:fim_analytics}

We derive the asymptotic prefactor
$C_\infty = \sqrt{g_{F}^\top\mathcal{M}^{-1}g_{F}}$ semi-analytically. For the grid sampling strategy, we provide a full analytic derivation below, while the derivations for the $W^2$ and $|W|$ strategies combine analytic approximations with numerical evaluation.
Four approximations are used: (i)~the continuum
limit $\delta\to 0$; (ii)~large-$\alpha_0$ separation, so cross-component
terms $\sim\mathcal{O}(e^{-2\alpha_0^2})$ are dropped; (iii)~rapid fringe
oscillation averaging, $\langle\cos^2 s\rangle = \langle\sin^2 s\rangle = \tfrac12$
(where $\langle\cdot\rangle$ denotes spatial averaging; since $s = 4\alpha_0 I$ leads to rapid
oscillation on the scale of the Gaussian width, cross-terms proportional
to $\langle\cos s\rangle$ or $\langle\sin s\rangle$ vanish and squared terms
average to $\tfrac12$);
and (iv)~evaluation at the symmetric even-cat point, where
$A_L = A_R \approx \tfrac12$, $A_C \approx \tfrac{1}{2}$,
all Gaussians are isotropic with $\sigma=\tfrac12$, and $\phi=\varphi=0$.
The Gaussians simplify to $\mathcal{N}_j(\beta)=\tfrac{2}{\pi}e^{-2|\beta-\mu_j|^2}$.

\paragraph{Wigner gradients.}\mbox{}
\\
\textit{Vanishing derivatives (4 parameters).}
With the assumptions above, the following four parameters have no impact on
the Wigner derivatives at the evaluation point: the three Gaussian rotation
angles $\theta_L$, $\theta_C$, $\theta_R$ and the fringe phase $\varphi$.
\begin{itemize}
\item $\partial_{\theta_j} W = 0$ for $j\in\{L,C,R\}$:
  at $\sigma^{(j)}_{x}=\sigma^{(j)}_{y}=\tfrac12$ the Gaussians are isotropic and
  the rotation-angle derivative vanishes identically.
\item $\partial_\varphi W = 0$:
  at $\varphi=0$ the fringe wavevector is aligned with the $I$-axis;
  its transverse derivative vanishes.
\end{itemize}
These four directions contribute neither Fisher information nor fidelity
gradient and play no role in $C_\infty$.

\textit{Non-vanishing derivatives (12 directions).}
\begin{align}
\label{eq:dW16_AL}
\frac{\partial W}{\partial A_L} &= \mathcal{N}_L - \mathcal{N}_R,
\\[2pt]
\label{eq:dW16_AC}
\frac{\partial W}{\partial A_C} &= -2\,\mathcal{N}_C\cos s,
\\[2pt]
\label{eq:dW16_ell_lobe}
\frac{\partial W}{\partial \ell_{L,x}} &= A_L\,\mathcal{N}_L\,(4\tilde I_L^2 - 1),
\quad
\frac{\partial W}{\partial \ell_{L,y}} = A_L\,\mathcal{N}_L\,(4\tilde Q_L^2 - 1),
\\[2pt]
\label{eq:dW16_ell_centre}
\frac{\partial W}{\partial \ell_{C,x}} &= -2A_C\cos s\cdot \mathcal{N}_C\,(4I^2 - 1),
\quad
\frac{\partial W}{\partial \ell_{C,y}} = -2A_C\cos s\cdot \mathcal{N}_C\,(4Q^2 - 1),
\\[2pt]
\label{eq:dW16_IL}
\frac{\partial W}{\partial I_L} &= 4A_L\tilde I_L\,\mathcal{N}_L - 4A_C I\,\mathcal{N}_C\cos s,
\\[2pt]
\label{eq:dW16_QL}
\frac{\partial W}{\partial Q_L} &= 4A_L\tilde Q_L\,\mathcal{N}_L - 4A_C Q\,\mathcal{N}_C\cos s
  - 4 I A_C \mathcal{N}_C\sin s,
\end{align}
where $\tilde I_L = I - I_L$, $\tilde Q_L = Q - Q_L$, and $s = 4\alpha_0 I$.
Here,
$\mu_L = (I_L, Q_L)$ and $\mu_R = (I_R, Q_R)$ are the mean positions of the left and the right Gaussians.
The right-Gaussian derivatives $\partial_{\ell_{R,k}} W$ and $\partial_{I_R,Q_R} W$
follow from left--right symmetry ($L \leftrightarrow R$ everywhere, with a sign
flip on the phase term in $\partial_{Q_R}W$).

In Eq.~\eqref{eq:dW16_AL}, the antisymmetric combination $\mathcal{N}_L-\mathcal{N}_R$ arises from
the normalisation constraint: varying $A_L$ simultaneously changes
$A_R = 1-A_L$ in the opposite direction.
In Eqs.~\eqref{eq:dW16_IL} and \eqref{eq:dW16_QL}, the terms proportional to
$\mathcal{N}_C$ come from the midpoint constraint $\mu_C = (\mu_L+\mu_R)/2$ moving the
central Gaussian.

\paragraph{Continuum FIM entries.}
Near the left Gaussian $L$, substituting into the continuum FIM formula and reducing to
component-centred polar coordinates, all entries reduce to the radial integral family
\begin{equation}
\label{eq:In_def}
I_n := \int_0^\infty \frac{r^{2n+1}\,e^{-4r^2}}{4-e^{-4r^2}}\,dr
      = \frac{n!\,L_{n+1}}{8\cdot 4^n},
\end{equation}
where $L_k := \mathrm{Li}_k(\tfrac{1}{4}) = \sum_{j=1}^\infty (1/4)^j/j^k$
are polylogarithm constants ($L_1 = \ln\tfrac43$, $L_2 \approx 0.268$,
$L_3\approx0.258$).
The key FIM entries at the evaluation point are
\begin{align}
\label{eq:M16_AA}
\mathcal{M}_{A_LA_L}
&= 2\pi L_1,
\\
\label{eq:M16_ell}
\mathcal{M}_{\ell_{j,k}\ell_{j,k}}
&= \frac{\pi}{4}\!\left(\frac{3}{4}L_3 - L_2 + L_1\right),
\\
\label{eq:M16_AC}
\mathcal{M}_{A_CA_C}
&= 2\pi L_1,
\\
\label{eq:M16_mumu}
\mathcal{M}_{I_LI_L}
&= \pi L_2 + \mathcal{O}\!\bigl((A_C/A_L)^2\bigr).
\end{align}
A parameter is a \textit{pure nuisance direction} if it enters the Fisher
matrix but contributes zero fidelity gradient, and therefore does not reduce
$C_\infty$.
The Gaussian cross-coupling from the $A_L$ nuisance direction has the opposite sign
on the two Gaussians:
$\mathcal{M}_{A_L,\ell_{Lx}} = m$ and
$\mathcal{M}_{A_L,\ell_{Rx}} = -m$
(with $m = \frac\pi2(\frac12 L_2 - L_1) < 0$),
a direct consequence of the antisymmetric derivative~\eqref{eq:dW16_AL}.
Position entries such as $\mathcal{M}_{I_LI_L}$ are nuisance directions
(parameters that contribute Fisher information but carry zero fidelity gradient,
and therefore increase $\mathcal{I}$ without reducing $C_\infty$; see below). \\

\paragraph{Fidelity gradient.}
The fidelity $F(\boldsymbol{\eta})$ (see Eq.~\eqref{eq:fidelity_def})
has the following gradient for a symmetric cat state 
\begin{equation}
\label{eq:gF16}
g_{F}
= \Bigl(
\underbrace{0,0,-\tfrac{1}{8},-\tfrac{1}{8},0}_{\rm left\;},\;
\underbrace{0,0,-\tfrac{1}{8},-\tfrac{1}{8},0}_{\rm right\;},\;
\underbrace{-\tfrac14,-\tfrac14,0}_{\rm centre},\;
\underbrace{0,-1,0}_{\rm amp./fringe}
\Bigr)^\top.
\end{equation}
The entries are given in the same order as the parameter vector $\boldsymbol{\eta}$; the
log-width components $-\tfrac18$ correspond to $\partial F/\partial\ell_{j,k}
= \sigma_{j,k}\,\partial F/\partial\sigma_{k}^{(j)}$.
The amplitude component vanishes exactly,
\begin{equation}
\label{eq:gF_AL_zero}
\frac{\partial F}{\partial A_L}\bigg|_{16}
= \pi\int W_{\rm cat}\,(\mathcal{N}_L - \mathcal{N}_R)\,d^2\beta
= \underbrace{\pi\!\int W_{\rm cat}\,\mathcal{N}_L\,d^2\beta}_{=\,1/2}
- \underbrace{\pi\!\int W_{\rm cat}\,\mathcal{N}_R\,d^2\beta}_{=\,1/2}
= 0,
\end{equation}
since both integrals equal $\tfrac12$ by the left--right symmetry of the
even-cat target.
Vanishing of the position gradients follows from the odd parity of
$\tilde I_j G_j$ and $\tilde Q_j G_j$ under inversion at $\mu_j$, together
with the $\mathcal{O}(\alpha_0^{-1})$ phase term in~\eqref{eq:dW16_QL}
averaging to zero against the slowly-varying $W_{\rm cat}$ envelope.
$A_L$ is a nuisance direction a direction: it produces the FIM entry
$\mathcal{M}_{A_LA_L} = 2\pi L_1$ (Eq.~\eqref{eq:M16_AA}) but
$\partial F/\partial A_L = 0$ by the symmetry argument above.
The remaining parameters with non-zero gradients, i.e. the six log-width parameters ($\ell_{j,k}$)
and $A_C$, form the \textit{signal sector}
$\boldsymbol{q} = (\ell_{L,x},\ell_{L,y},\ell_{R,x},\ell_{R,y},\ell_{C,x},\ell_{C,y},A_C)$,
with gradient $g_q = (-\tfrac18,-\tfrac18,-\tfrac18,-\tfrac18,
-\tfrac14,-\tfrac14,-1)^\top$.

\paragraph{Asymptotic prefactor: grid strategy.}
The signal-block CRB
$C_\infty = \sqrt{g_q^\top\mathcal{M}_{\rm eff}^{-1}g_q}$ decomposes into two independent
contributions. Here, $\mathcal{M}_{\rm eff} = \mathcal{M}_{qq} - \mathcal{M}_{qn}\mathcal{M}_{nn}^{-1}\mathcal{M}_{nq}$
is the Schur-complement effective FIM.

\textit{Gaussian block.}
The $4\times4$ effective block for $(\ell_{Lx},\ell_{Ly},\ell_{Rx},\ell_{Ry})$
receives a Schur correction from the $A_L$ nuisance direction.
Because the $A_L$--width coupling has \emph{opposite signs} on the two Gaussians
($+m$ left, $-m$ right), the cross-block correction is positive
($c = +m^2/\mathcal{M}_{A_LA_L} > 0$).
A block-diagonalisation of the $4\times4$ matrix shows that this correction
cancels exactly in the CRB projection, giving the closed-form result
\begin{equation}
\label{eq:Cinf_lobe_app}
C_{\infty,\rm Gaussian}^{2}
= \frac{1}{4\pi(2L_1 - 2L_2 + L_3)}.
\end{equation}

\textit{Centre block.}
The $3\times3$ block for $(\ell_{Cx},\ell_{Cy},A_C)$ has a single finite
eigenvector $\hat{v}_2 = (1,1,-1)/\sqrt{3}$ that couples to $g_{\rm center} = (-\tfrac14,-\tfrac14,-1)$
with $(g_{\rm center}^\top\hat{v}_2)^2 = \tfrac{1}{12}$ (exact: $(-\tfrac14-\tfrac14+1)/\sqrt{3} = 1/(2\sqrt{3})$).
Its eigenvalue is
$\lambda_2 = \tfrac\pi6\,\mathcal{S}_3$,
where
\begin{equation}
\label{eq:S3_def}
\mathcal{S}_3 := \sum_{n=1}^\infty \frac{\binom{2n}{n}}{4^n\,n^3} \approx 0.5678
\end{equation}
is a central-binomial sum.  This gives
$C_{\infty,\rm center}^{2} = 1/(2\pi\mathcal{S}_3)$.

\textit{Full result (grid sampling).}
Combining both blocks:
\begin{equation}
\label{eq:Cinf_full_app}
\boxed{
C_{\infty,\rm grid}
= \sqrt{\frac{1}{4\pi(2L_1-2L_2+L_3)}
       + \frac{1}{2\pi\,\mathcal{S}_3}}
\approx 0.74.
}
\end{equation}

\paragraph{Asymptotic prefactor: $W^2$-optimal strategy.}
For $W^2$-optimal sampling, the measurement budget is concentrated at pixels
proportional to $W_{\rm cat}(\beta)^2$, i.e.\ $p(\beta) \propto
W_{\rm cat}(\beta)^2$.

The $W^2$ weight introduces an extra Gaussian factor $W_L^2 \propto e^{-4r^2}$
inside the integral, defining a new radial integral family
\begin{equation}
\label{eq:Jn_def}
J_n := \int_0^\infty \frac{r^{2n+1}\,e^{-8r^2}}{4 - e^{-4r^2}}\,dr
     = \frac{n!}{32\cdot 4^n\,\gamma^2}\bigl[\mathrm{Li}_{n+1}(\gamma)-\gamma\bigr]
       \bigg|_{\gamma=1/4},
\end{equation}
with values $J_0\approx 0.0754$, $J_1\approx 0.00878$, $J_2\approx 0.000563$.
Replacing $I_n\to J_n$ throughout, the weighted FIM Gaussian entries become
\begin{align}
\label{eq:Mhat_AA}
\mathcal{M}_{A_LA_L} &= 16\pi J_0,
\\
\label{eq:Mhat_ell}
\mathcal{M}_{\ell_{j,k}\ell_{j,k}}
  &= 8\pi A_L^4\bigl(6J_2 - 4J_1 + J_0\bigr),
\\
\label{eq:Mhat_Aell}
\mathcal{M}_{A_L,\ell_{L,x}}
  &= m,\quad
\mathcal{M}_{A_L,\ell_{R,x}} = -m,\quad
m = 8\pi A_L^3(2J_1-J_0).
\end{align}
The $A_L$--width coupling retains opposite signs on the two Gaussians; by the same
block-diagonalisation argument as above the Schur correction cancels exactly in
the CRB projection, yielding the Gaussian-block closed form
\begin{equation}
\label{eq:Cinf_lobe_W2}
C_{\infty, W^2,\rm Gaussian}^{2}
= \frac{1}{4\cdot 8\pi A_L^4(2J_2 - 2J_1 + J_0)}.
\end{equation}

The full $C_{\infty,W^2}$ requires numerical evaluation of the complete
signal-block FIM with the exact cat Wigner function:
\begin{equation}
\label{eq:Cinf_W2_16_value}
\boxed{C_{\infty,W^2} \approx 1.76.}
\end{equation}

\paragraph{Asymptotic prefactor: $|W|$-optimal strategy.}
For $|W|$-optimal sampling, the measurement budget is concentrated at pixels
proportional to $|W_{\rm cat}(\beta)|$, i.e.\ $p(\beta) \propto
|W_{\rm cat}(\beta)|$.

The $|W|$ weight introduces an intermediate Gaussian factor $|W_L|\propto
e^{-2r^2}$, defining
\begin{equation}
\label{eq:Kn_def}
K_n := \int_0^\infty \frac{r^{2n+1}\,e^{-6r^2}}{4 - e^{-4r^2}}\,dr
     = \frac{n!}{32\cdot 4^n}\,\Phi\!\bigl(\tfrac14,\,n+1,\,\tfrac32\bigr),
\end{equation}
where $\Phi(z,s,a):=\sum_{k=0}^\infty z^k/(k+a)^s$ is the Lerch transcendent.
Numerically $K_0\approx 0.0249$, $K_1\approx 0.00194$, $K_2\approx 0.000181$.
Replacing $J_n\to K_n$ and $A_L^4\to A_L^3$, the weighted FIM Gaussian entries are
\begin{align}
\label{eq:Mtilde_AA}
\mathcal{M}_{A_LA_L} &= 16\pi K_0,
\\
\label{eq:Mtilde_ell}
\mathcal{M}_{\ell_{j,k}\ell_{j,k}}
  &= 8\pi A_L^3\bigl(6K_2 - 4K_1 + K_0\bigr),
\\
\label{eq:Mtilde_Aell}
\mathcal{M}_{A_L,\ell_{Lx}}
  &= m,\quad
\mathcal{M}_{A_L,\ell_{Rx}} = -m,\quad
m = 8\pi A_L^2(2K_1-K_0).
\end{align}
The Schur correction again cancels in the CRB projection:
\begin{equation}
\label{eq:Cinf_lobe_absW}
C_{\infty,|W|, \rm Gaussian}^{2}
= \frac{1}{4\cdot 8\pi A_L^3(2K_2 - 2K_1 + K_0)}.
\end{equation}
As in the $W^2$ case, the full $C_{\infty,|W|}$ requires numerical
evaluation:
\begin{equation}
\label{eq:Cinf_absW_16_value}
\boxed{C_{\infty,|W|} \approx 2.07.}
\end{equation}

The asymptotic prefactors are summarised in Table~\ref{tab:crb_prefactors}. These bounds are verified numerically and plotted against the total measurement budget $N_{\rm total}$ and the cat state size $\alpha_0$ in Fig.~\ref{fig:uncertainity_sampling_strategies}.
\begin{table}[htbp]
\centering
\begin{tabular}{lcc}
\hline
Strategy & $C_\infty$ & Normalisation \\
\hline
Grid  & 0.74 & $\sigma_F\sqrt{N_{\rm shots}}/\delta$ \\
$W^2$ optimal & 1.76 & $\sigma_F\sqrt{N_{\rm total}}$ \\
$|W|$ optimal & 2.07 & $\sigma_F\sqrt{N_{\rm total}}$ \\
\hline
\end{tabular}
\caption{Asymptotic CRB prefactors $C_\infty$ for the constrained
16-parameter model.
The grid value is the exact analytic result of
Eq.~\eqref{eq:Cinf_full_app}; the optimal strategy values are obtained by
numerical evaluation of the full signal-block FIM with the exact cat Wigner function.}
\label{tab:crb_prefactors}
\end{table}

\subsection{Comparison of fidelity uncertainty lower bound for different sampling strategies}
\label{sec:fim_comparison}
We compare the lower bound on the fidelity uncertainty $\sigma_F$ (Eq.~\eqref{eq:crb_fidelity}) as a function of the total measurement budget $N_{\rm total}$ and the cat state amplitude $\alpha_0$ for the four sampling strategies considered in this work: grid sampling, $W^2$ sampling, $|W|$ sampling, and adaptive sampling. The $|W|$ sampling strategy assumes single-shot parity measurements ($N_{\rm shots}=1$), whereas the adaptive, grid, and $W^2$ strategies use $N_{\rm shots}=40$ single-shot measurements per phase-space point. For adaptive sampling, the measurement points are generated from QuTiP simulations using the reconstruction algorithm with the hyperparameters specified in Sec.~\ref{app:OptimalParameters}. Since the simulated single-shot measurements are less noisy than the experimental data, we set the measurement noise parameter in Eq.~\ref{eq:sample_likelihood_detailed} to $\sigma = 0.04$.

\begin{figure}[!tbp]
\centering
\includegraphics[width=\linewidth]{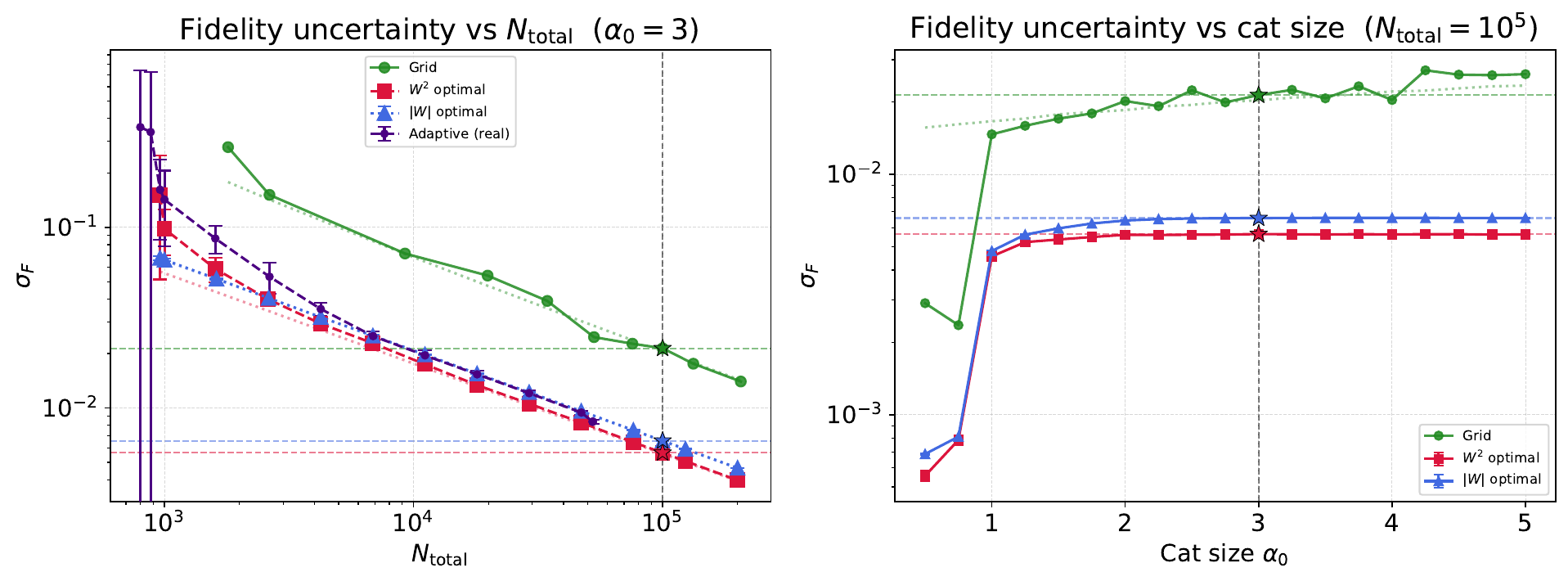}
\caption{
    Fidelity uncertainty $\sigma_F$ as a function of the total measurement budget $N_{\rm total}$ (left) and cat state amplitude $\alpha_0$ (right). For the $W^2$ and $|W|$ sampling strategies, each data point is the mean over 10 independent Monte Carlo runs (error bars indicate the run-to-run standard deviation), with the Fisher information matrix evaluated from the 16-parameter Wigner function gradient (Sec.~\ref{sec:fim_analytics}) at the sampled phase-space points. For the adaptive sampling strategy, each point is the mean over 10 independent simulated reconstruction runs.
    \textbf{Left}: $\sigma_F$ as a function of $N_{\rm total}$ for $\alpha_0=3$ (log--log scale). The grid strategy ($N_{\rm shots}=40$) employs a deterministic $n_I\times n_Q$ grid with equal spacing along both phase-space axes (ranging from $9\times5$ to $101\times51$ pixels for $\alpha_0=3$). Dotted lines indicate the asymptotic scalings $\sigma_F=C_\infty/\sqrt{N_{\rm total}}$ for the Monte Carlo strategies and $\sigma_F=C_{\infty,\mathrm{grid}}\,\delta/\sqrt{N_\textrm{shots}}$ for the grid strategy (Eq.~\eqref{eq:Cinf_def}).
    \textbf{Right}: $\sigma_F$ as a function of $\alpha_0$ for a fixed measurement budget $N_{\rm total}=10^5$. For the grid sampling, $n_I$ is adjusted for each $\alpha_0$ such that $n_I\times n_Q\approx10^5$, ensuring an equal measurement budget across all strategies. The green dotted line shows the theoretical prediction $\sigma_F = C_{\infty,\mathrm{grid}}\,\delta(\alpha_0)/\sqrt{N_{\rm shots}}$, where $\delta(\alpha_0)=2(\alpha_0+3)/(n_I-1)$ grows with cat size to capture fully the Wigner function. 
    \\
    In both panels, stars indicate the reference point $(\alpha_0,N_{\rm total})=(3,10^5)$. At this operating point, optimal sampling strategies reduce $\sigma_F$ by more than a factor of $2$ compared with grid sampling, highlighting the advantage of concentrating measurements in the most informative regions of phase space. Note that the experimental results in Sec.~\ref{sec:SamlingComparison} use a maximum measurement budget of $N_{\rm total}=2\times10^5$, where the factor of two arises from the background Wigner measurement.
}
\label{fig:uncertainity_sampling_strategies}
\end{figure}

\begin{figure}
    \centering
    \includegraphics[width = \linewidth]{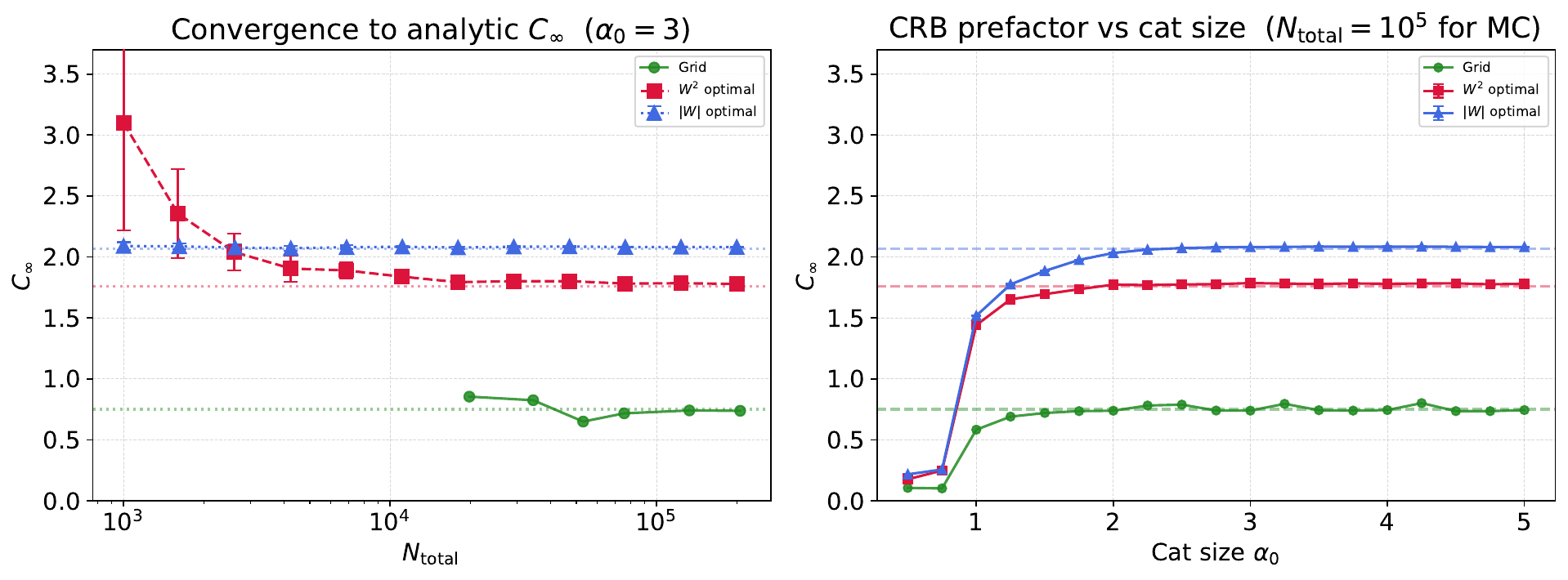}
    \caption{Convergence to the asymptotic prefactor $C_\infty$ for different sampling strategies obtained by the same settings of Fig.~\ref{fig:uncertainity_sampling_strategies}. Dashed lines indicate the asymptotic value $C_\infty$ for each sampling strategy. For the $W^2$ and $|W|$ sampling strategies, each data point is the mean over 10 independent Monte Carlo runs (error bars indicate the run-to-run standard deviation).  Dotted horizontal lines indicate the analytically derived asymptotic values $C_{\infty,W^2} \approx 1.76$ and $C_{\infty,|W|} \approx 2.07$ (Table~\ref{tab:crb_prefactors}).
For the grid sampling, $C_\infty = \sigma_F \sqrt{N_{\rm shots}} / \delta$ converges to the analytic asymptote $C_{\infty,\rm grid} \approx 0.74$ (dashed line) as $\delta \to 0$. Note that a comparison of $C_\infty$ is meaningful only between sampling strategies that share the same normalisation. In particular, the grid normalisation, $\sigma_F\sqrt{N_{\rm shots}}/\delta$, and the optimal strategy normalisation, $\sigma_F\sqrt{N_{\rm total}}$, cannot be compared directly.
\textbf{Left}: all three strategies converge to their respective $C_\infty$ values as $N_{\rm total}$ increases, reaching their asymptotes for $N_{\rm total} \gtrsim 10^4$.
\textbf{Right}: all strategy prefactors are nearly flat across all cat sizes $\alpha_0 \in [2, 5]$, converging to $C_\infty$.
}
    \label{fig:Cinf_convergence}
\end{figure}

The results are presented in Fig.~\ref{fig:uncertainity_sampling_strategies}. As expected, the optimal and adaptive sampling strategies consistently outperform the grid sampling, demonstrating the benefit of concentrating measurements in the most informative regions of phase space. In the asymptotic limit of a large number of measurements, the adaptive sampling strategy exhibits a convergence rate comparable to that of $|W|$ sampling: fitting the linear part of the adaptive sampling convergence curve $(N > 10^4)$ gives $C_\infty = 2.00 \pm 0.03$, which is close to a $|W|$ sampling value $C_\infty = 2.07$. Both optimal sampling strategies approach the analytically predicted scaling, $C_{\infty}/\sqrt{N_\textrm{total}}$, for $N_{\rm total}\gtrsim10^4$, as shown in Fig.~\ref{fig:Cinf_convergence}. For a low number of measurements, grid sampling becomes biased since the phase space resolution is insufficient to resolve the Wigner function features. Consequently, these points are omitted from the figure. The first data point shown for the grid corresponds to a $9\times5$ sampling grid for the $\alpha_0=3$ cat state.
    
\clearpage

\bibliography{cat_reconstruction}

\end{document}